\documentclass[prd,superscriptaddress,onecolumn,showpacs,11pt,msmath,
preprintnumbers,showkeys]{revtex4}
\usepackage{wrapfig,rotating}
\usepackage{graphicx,epsfig,color}

\makeatletter

\newenvironment{figurehere}
{\def\@captype{figure}}
{}
\makeatother

\usepackage{graphicx}
\usepackage{dcolumn}
\usepackage{bm}

\def\beq{\begin{equation}}
\def\eeq{\end{equation}}
\def\beeq{\begin{eqnarray}}
\def\eeeq{\end{eqnarray}}

\def\2GPD{$_2\mbox{GPD}$}

\def\12{$1\otimes 2$}
\def\22{$2 \otimes 2$}

\def\Qsep{Q_{\mbox{\rm\scriptsize sep}}}
\def\Qsep2{Q^2_{\mbox{\rm\scriptsize sep}}}

\begin{document}

\title{The three point asymmetric cumulants in high multiplicity pp collisions.}

\author{B.\ Blok,
 R.\ Segev
\\[2mm] \normalsize Department of Physics, Technion -- Israel Institute of Technology,
Haifa, Israel}

\begin{abstract}
We study the influence of quantum interference and colour flow on three point correlations  described by asymmetric cumulants 
in high multiplicity events in pp collisions. We use the model previously developed for the study of
the collectivity in symmetric cumulants. We show that the resulting three point asymmetric cumulant 
is in qualitative agreement with the 
experimental data for the same parameters of the model as it was with the symmetric cumulants.
Our results show that the initial state correlations must play a major role and may be even dominant 
in the explanation of the correlations in high multiplicity pp events.
 \end{abstract}

  \maketitle
\thispagestyle{empty}

\vfill

\section{Introduction.}

\par Sizeable n-th harmonic coefficients $v_n$ for azimuthal momentum asymmetries have been observed at the LHC in nucleus-nucleus (AA), proton-nucleus (pA) and proton-proton (pp) collisions  \cite{1,2,3,4,5,6}. These asymmetries  indicate a collective mechanism that relates all particles produced in a given collision. The dynamical origin of these collectivity phenomena continues to be sought in competing and potentially contradicting pictures.
\par There are two basic approaches towards the explanation of this collective behaviour. The first approach is based on the final state interactions, like viscious fluid dynamics simulations \cite{7} or kinetic transport models 
\cite{8,9,10,11,11a} of heavy ion collisions. In the AA collisions the jet quenching phenomena provides an alternative confirmation for  such a approach. However the jet quenching is missing in smaller
pp and pA collision systems. Moreover, in marked contrast to any final state explanation of flow anisotropies $v_n$ in pp collisions, the phenomenologically successful modelling of soft multi-particle production in modern multi-purpose pp event generators \cite{12} are based on free-streaming partonic final state distributions supplemented by independent fragmentation into hadrons. Efforts to go beyond this picture are relatively recent, see e.g. \cite{13,14}. Therefore two contradictory pictures to describe the multiparticle dynamics in small systems currently exist-- the one based on the  final state interactions, and the second that does not involve the final state interactions. One approach in the second direction corresponds to the  recent works in the framework of Colour Glass Condensate (CGC) \cite{14a,14b,14c}, see \cite{14d} for a recent review, based on the parton saturation hypothesis, that recently made significant progress towards phenomenological description of correlations in small systems.

\par Recently  a QCD based simple model, based on the theory of multiparton interactions (MPI) in pp collisions
\cite{16a,16b,16c,16d,16e,16g,16k,16l,16n,16m} , and not involving the saturation effects,
 was proposed in \cite{15,16} 
 to study the effects of quantum interference and colour flow in high multiplicity pp events.
 The strong simplification of the model consists in neglecting a dynamically explicit formulation of the
scattering process: all gluons in the incoming wave function are assumed to be freed in the scattering process with the same (possibly
small) probability. The model pictures the incoming hadronic wave function as a collection of N colour sources in adjoint representation
distributed in transverse space according to a classical density $\rho(\vec r_i)$. On the amplitude level, emission of a gluon is   taken into account in soft gluon/eikonal approximation.
\par  In Ref. \cite{16}  the flow coefficients $v_n\left\{2s\right\}$, determined by $2s$ point symmetric cumulants 
\beq
sc_{n}\left\{ 2s\right\} \equiv\left\langle \left\langle e^{in(\sum_{i=1}^{i=s} \phi_i-\sum_{i=s+1}^{i=2s}\phi_i)}\right\rangle \right\rangle
\label{0}
\eeq
were calculated. Here $<<>>$ means averaging over the multiparticle final states and 
taking the cumulant. The phases $\phi_i$ are the azimuthal angles of measured soft hadrons.

\par The model \cite{15,16}    predicts both the collectivity phenomena  and  the qualitatively correct scale of correlations as well as their behaviour as the functions of the transverse momenta.
\par Consequently,  it makes sense to study the other recently measured flow phenomena in high multiplicity pp collisions
in this  framework of this model.

 One group of potentially interesting cumulants  are the three point  asymmetric cumulants,
The corresponding cumulants are often denoted as $ac_n\left\{ 3\right\} $:
\beq
ac_{n}\left\{ 3\right\} \equiv\left\langle \left\langle e^{in\left(\phi_{1}-\phi_{3}\right)}e^{in\left(\phi_{2}-\phi_{3}\right)}\right\rangle \right\rangle =\left\langle \left\langle e^{in\left(\phi_{1}+\phi_{2}-2\phi_{3}\right)}\right\rangle \right\rangle .
\label{as1}
\eeq
These cumulants were recently studied experimentally \cite{17}. Since we consider in this paper only $ac_n\left\{3\right\}$ cumulants,
we shall denote them as simply $ac_n$ below.
\par The purpose of the paper is to calculate the three point  correlations (\ref{as1}) and to compare these correlators to  the available experimental data \cite{17}. We shall see that our results are in qualitative agreement with the experimental data although there is not enough experimental data for detailed comparison. 
\par We shall give the detailed predictions for transverse momenta dependence and for the scale
(characteristic magnitude) of the three point cumulant .
We shall also discuss the dependence of this cumulant on multiplicity. 
\par Recall that the model \cite{15,16} was based on large $N_c$ and $N$ expansion and then the results were extrapolated to $N_c=3$.
\par The paper is organized in the following way. In the second chapter we describe the model and formalism, in the third chapter we consider the simplest case of N=m=3-three sources and 3 gluons. We calculate the basic diagrams contributing to
cumulant.  In the fourth chapter we sum the whole leading in $m^2/(N^2_c-1)$ contribution, which includes configurations due to one "tripole"  or "dipole" with 3 gluons and unintersecting dipoles. We find the multiplicity dependence of our results. In section 5 we discuss the relation of our calculations to the experimental data
for correlator (\ref{as1}). The  conclusions are presented in section 6.

\section{Basic formalism.}
\subsection{The Model \cite{15,16}}
\par Each $pp$ collision is considered as an event consisting of $N$ emitting sources characterized by 
 two-dimentional transverse positions $\boldsymbol{r}_{j}$
and the initial colours in the adjoint representation $b_{j}$. 
Physically these sources correspond to multiparton interactions (MPI). In other words each source is a collision of two
partons-one from  each of the colliding nucleons.

The emission amplitude of the gluon with colour a  and transverse momentum $\vec k$  from a source in transverse position $\vec r _j$
 is given by an eikonal vertex:
\beq
T_{b_{j}c_{j}}^{a}\overrightarrow{f}\left(\boldsymbol{k}\right)e^{i\boldsymbol{k}\cdot\boldsymbol{r}_{j}},
\eeq
where $T_{b_{j}c_{j}}^{a}$ are the adjoint generators of
 $SU\left(N_c\right)$, and $\overrightarrow{f}\left(\boldsymbol{k}\right)$ is a vertex function. The concrete form of $\vec f$ will not influence  our results
(see below). For example, for the case of coulombic radiation we have $\vec f=\vec k/k^2$. However since the relevant momenta are small the function f must be taken   to be a nonperturbative one. In the cross section the emitted gluon can be absorbed by the same source 
in the complex conjugated amplitude ("diagonal gluons") or by different source ("off-diagonal gluons"), leading to 
multiparticle correlations.The simplest diagrams contributing to multiparticle cross section and to correlations are presented in Fig.1. In the left there is a diagram with 2 sources and 2 diagonal gluons, in the right there are still 2 sources but gluons are off-diagonal:they are emitted by one source and absorbed by another leading to azimuthal correlations.
  \par After   calculating cross sections for given source positions we average over  the source positions with a classical probability distribution $\rho\left(\left\{ \boldsymbol{r}_{j}\right\} \right)$, corresponding to the 
  distribution of multiparton interactions in the pp collision \cite{16b}:
\beq
  \frac{d\hat \sigma}{d^2\vec k_1....d^2\vec k_m}=\prod_{i=1}^N\int d^2 \boldsymbol{r}_i\rho(\boldsymbol{r}_i)\hat \sigma(\vec k_i,\boldsymbol{r}_i),
  \eeq
  where $\sigma(k_i,\boldsymbol{r}_i)$ is the cross section of production of m gluons for sources in fixed
   transverse positions $ \boldsymbol{r}_i$.

  In this paper we shall neglect so called $1\rightarrow 2$ mechanism for MPI  \cite{16b,16e,16k}., and carry all calculations in the mean field approximation. In this case the source/MPI distribution  in pp system has a gaussian form:
\beq
\rho\left(\left\{ \boldsymbol{r}_{j}\right\} \right)=\prod_{i=1}^{i=N}\frac{1}{2\pi B}e^{\frac{-r_{i}^{2}}{2B}},
\eeq
where the parameter B is determined from the analysis of the one particle GPD data at HERA \cite{18}.
As it was noted in \cite{15}, the mean field approach to MPI corresponds to $B=4$ GeV$^{-2}$,
the actual experimental data reparametrized in the mean field form, i.e. assuming factorization of MPI
cross-sections corresponds to $B=2$ GeV$^{-2}$, and the best fit for experimental data for symmetric correlators 
$sc_n$ considered in \cite{15,16} corresponds to $B=1$ GeV$^{-2}$, this was justified (though 
not proved) in \cite{15} by arguments due to possible contribution of very small dipoles due to so called $1\rightarrow 2$
mechanism in MPI \cite{16b,16e,16k}. In this paper we shall see that the value of B influences only transverse momentum dependence, and the value $B=1$ GeV$^{-2}$ seems to be in the best agreement with the experimental data.
\par We shall work in the limit of a large number of sources $N$, $m$ finite, $N\rightarrow\infty$ \cite{15,16},
and $N_c\rightarrow\infty$ and classify all diagrams in powers of $1/N_c,1/N$.
\begin{figure}[htbp]
\hskip 6cm
\includegraphics[scale=0.65]{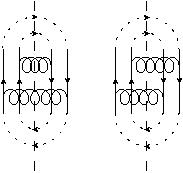}\hspace{6cm}
\label{Fig1}
\caption{The simplest   diagrams contributing to total cross section and to correlations: left-2 diagonal gluons, right-interference corresponding to two off-diagonal gluons forming dipole.}
\end{figure}
\subsection{The Correlation Functions.}
The correlation functions we are interested in have the general form 
\beq
K^{(n)}(k_1,..,k_s)=M_s\frac{\int_{\rho} d\phi_1...d\phi_s\exp{[i(\sum_{i=1}^{i=s}n_i\phi_i)]}\frac{d^sN}{d\Gamma_1...d\Gamma_s}}{(2\pi)^{s}\prod_{i=1}^{i=s}\int_{\rho}\frac{dN}{d\Gamma_i}},
\eeq
where $\sum_i n_i=0$, and $\int_{\rho}=\int  \prod_i d^2y_i\rho(\vec y_i)$ is the averaging over position of the sources.
The standard s-particle spectrum  has the form
\beq
\frac{d^sN}{d\Gamma_1...\Gamma_s}={m\choose s}\frac{d^s\hat\sigma}{\hat\sigma d\Gamma_1...d\Gamma_s}.
\eeq
Here $\hat\sigma$ is the cross section for the production of s gluons, and the normalisation factor $M$ is fixed as 
\beq  M_s=m^s/{m\choose s}.
\label{norm}
\eeq
Here $m$  is the total multiplicity, and s is the number of measured gluons. We assume local parton-hadron  duality  (LPHD)\cite{18}, so 
 radiated gluon correlations and multiplicities coincide with correlations and  multiplicities in the soft hadronic spectrum,
 The number of radiated gluons is in one to one correspondence with a number of radiated soft hadrons.
\par In this paper we shall be interested in the case $s=3,n_1=n,n_2=n,n_3=-2n$,i.e.  in 
\beq
K^{(n)}(k_1,k_2,k_3)=M_3\frac{\int_{\rho} d\phi_1...d\phi_3\exp{[in(\phi_1+\phi_2-2\phi_3)]}\frac{d^3N}{d\Gamma_1...\Gamma_3}}{(2\pi)^s\prod_{i=1}^{i=3}\int_{\rho}\frac{dN}{d\Gamma_i}}.
\label{as}
\eeq
\subsection{The differential cross section.}
The relevant differential cross section was calculated in \cite{15}:
We can now write $\hat{\sigma}$ explicitly in the limit $N\rightarrow \infty$ , i.e. omitting terms that are zero as $N\rightarrow\infty$
\begin{eqnarray}
\hat{\sigma}&\propto& N_{c}^{m}(N_{c}^{2}-1)^{N}N^{m}\prod^{m}_{i=1}|\overrightarrow{f}\left(\boldsymbol{k}_{i}\right)|^{2}\nonumber\\[10pt]
&\times&(1+\frac{F_{corr}^{(2)}(N,m)}{N^{2}(N_{c}^{2}-1)}{\sum_{ab}}{\sum_{lm}}2^{2}cos\left(\mathbf{k}_{a}\cdot\Delta\mathbf{r}_{lm}\right)cos\left(\mathbf{k}_{b}\cdot\Delta\mathbf{r}_{ml}\right)\nonumber\\[10pt]
&+&\frac{F_{corr}^{(3)}\left(N,m\right)}{4N^{3}\left(N_{c}^{2}-1\right)}{\sum_{(abc)}}{\sum_{(lm)}}2^{3}\cos\left(\mathbf{k}_{a}\cdot\Delta\mathbf{r}_{lm}\right)\cos\left(\mathbf{k}_{b}\cdot\Delta\mathbf{r}_{lm}\right)\cos\left(\mathbf{k}_{c}\cdot\Delta\mathbf{r}_{lm}\right)\nonumber\\[10pt]
&+&\frac{F_{corr}^{(3)}\left(N,m\right)}{N^{3}\left(N_{c}^{2}-1\right)^{2}}{\sum_{(abc)}}{\sum_{(lm)(mn)}}2^{3}\cos\left(\mathbf{k}_{a}\cdot\Delta\mathbf{r}_{lm}\right)\cos\left(\mathbf{k}_{b}\cdot\Delta\mathbf{r}_{mn}\right)\cos\left(\mathbf{k}_{c}\cdot\Delta\mathbf{r}_{nl}\right)\nonumber\\[10pt]
&+&O\left(N^{-4}\right)).\nonumber\\[10pt]
\label{3a}
\end{eqnarray}
Here the sums go over all ordered combinations of off-diagonal gluons and non ordered combinations of sources. Using 
Eq.(\ref{3a}) we can find any correlation function for any number of particles.
\par The first term in Eq.(\ref{3a}) corresponds to diagonal gluons, the second is a dipole term, which is a leading contribution to symmetric cumulants \cite{16}. 
The leading contribution to asymmetric correlator (\ref{as1}) comes from the third and fourth terms in the expansion (\ref{3a}) corresponding to 3-gluon dipole (which is actually 1/N suppressed relative to a second term) and a tripole diagram, depicted in the diagram of Fig. 2. Note that the tripole diagram has a finite value in
 the $N\rightarrow \infty$ limit.
 \par The factors $F_{corr}^{(2)}(N,m),F_{corr}^{(3)}(N,m)$ correspond to the contribution of diagonal gluons to
 the interference diagrams. As it was shown in  \cite{15} the diagonal gluons lead to the multiplicative renormalisation 
 of the correlators, given by the corresponding coefficients $F_{corr}$.
\subsection{The expansion.}
\par Recall the structure of the expansion discussed in \cite{15,16} for even harmonics for symmetric cumulants. 
The expansion was in two parameters: $1/(N_c^2-1)$ and in $1/N$. The leading contribution came from the dipole
diagram which was of order $1/(N^2_c-1)$, and the leading approximation in $1/(N^2_c-1),1/N$ was considered , so that
we could discard all $1/N$ suppressed diagrams. (Note however that odd harmonics appeared only due to  1/N suppressed
terms in the differential cross section expansion). It was shown in \cite{15}, that 
 the real parameters of the expansion for given multiplicity were $m^2/((N^2_c-1)N)$ where N  is the number of sources, and m is the multiplicity. The leading terms in this expansion however can be resummed, as it was done in \cite{16}. The corresponding diagrams are built from up to  $[N/2]$  sources  and correspond to nonintersecting dipoles ($[]$ means the integer part).
\par For the case of the 3 point cumulant the situation  is more complicated. For the symmetric cumulants we were 
able to show that the $1/N$ suppressed diagrams can be neglected. On the other hand for the three point cumulant 
the leading diagram in the $N\rightarrow \infty$ limit is a tripole (see Fig. 3a). However this diagram is suppressed 
by $1/(N^2_c-1)^2$ in the large $N_c$ limit. On the other hand the diagram corresponding to  the dipole with three off-diagonal 
gluons (Fig. 3b)  has zero limit for large N, i.e. it is suppressed as $1/N$, while it is only $1/(N^2_c-1)$ in the large $N_c$ limit.
Thus we shall expect that 1/N corrections will play a significant role for three point correlator  if we extrapolate to finite $1/N_c$.
In fact we shall see below that the dipole with 3 gluons will dominate numerically up to rather large number of sources.
 Thus we shall  take into account both the tripole and the dipole with three off-diagonal gluons. We shall also take into account
 the leading terms in the expansion in $m^2/(N^2_c-1)$ for both leading and subleading (1/N, i.e. dipole with 3 off-diagonal gluons) terms in the expansion. It is easy to show in analogy to the symmetric case \cite{16} that such expansion
 corresponds to the inclusion of up to $[N/2]$ nonintersecting integrated out dipoles with  2 off-diagonal gluons. 
 (By "integrated out" we mean that we integrate over momenta of the corresponding nonobservable off-diagonal gluons).
 It is easy to see that all other diagrams are subleading, i.e. suppressed  by higher powers of $1/(N^2_c-1)$, or $1/N$,
 i.e as $1/(N_c^2-1)^a1/N^b,a+b\ge 3$.
 \par For numerical calculations we extrapolate our results to finite $N_c=3$ and to finite N, actually fixing 
 $\bar m =m/N$, in the limit $m,N\rightarrow \infty$ as the parameter of the  model. We use the same values of $\bar m$ as it was done for symmetric cumulants in \cite{16}.

\section{The Tripole and 3-gluon dipole.}
\begin{figure}[htbp]
\begin{center}
\includegraphics[scale=0.65]{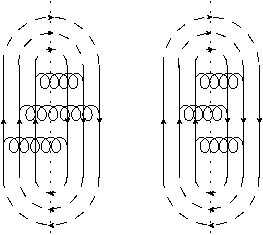}\hspace{0.1cm}
\label{Fig2}
\caption{Left: The tripole diagram for $N=m=3$. Right: 3-gluon dipole diagrams for $N=m=3$.}
\end{center}
\end{figure}

\par In this section we shall consider the simplest case of N=m=3. There are two contributions in this case:
first from the tripole diagram, second from dipole with 3 gluons. Note that each of these two diagrams will be 
the building block for the case of arbitrary m, N.
\par We shall start from analysing the single tripole term, that corresponds to the case $N=m=3$. The corresponding 
diagram is depicted in Fig. 2(left).
Note that in this case  there are 3!=6 ordered combinations of gluons, in addition there are 3 ways to put the phases 
on 3 available gluons, so we get, using Eq. (\ref{3a})  the multiplier 144:
 
\begin{eqnarray}
ac^{(3,3)}_n&\equiv&\frac{T_n}{(N^2_c-1)^2}\nonumber\\[10pt]
&=&\frac{144}{(N^2_c-1)^23^3}\int_{\rho}\int d\phi_{1}d\phi_{2}d\phi_{3}e^{in\left(\phi_{1}+\phi_{2}-2\phi_{3}\right)}\cos\left(\mathbf{k}_{1}\cdot\Delta\mathbf{r}_{12}\right)\cos\left(\mathbf{k}_{2}\cdot\Delta\mathbf{r}_{23}\right)\cos\left(\mathbf{k}_{3}\cdot\Delta\mathbf{r}_{31}\right),\nonumber\\[10pt]
\label{dt}
\end{eqnarray}
where $ac^{3,3} $ is the contribution of the tripole diagram into the total $ac$.
\par Consider now  the contribution of 3-gluon dipole to the asymmetric three point cumulant. The corresponding 
diagram is depicted in Fig. 2 (right).
We note that for the 3-gluon dipole only two of the 3 sources are involved and so the multiplier gets a factor of $\left(\begin{array}{c}3\\2\end{array}\right)$ and a factor of $4^{-1}$ from the colour trace, and the multiplier becomes 108:
\begin{eqnarray}
ac^{(3,2)}_n&\equiv&\frac{\tilde{T}_n}{N^2_c-1}\nonumber\\[10pt]
&=&\frac{144{3\choose 2}/4}{(N^2_c-1)3^3}\int_{\rho}\int d\phi_{1}d\phi_{2}d\phi_{3}e^{in\left(\phi_{1}+\phi_{2}-2\phi_{3}\right)}\cos\left(\mathbf{k}_{1}\cdot\Delta\mathbf{r}_{12}\right)\cos\left(\mathbf{k}_{2}\cdot\Delta\mathbf{r}_{12}\right)cos\left(\mathbf{k}_{3}\cdot\Delta\mathbf{r}_{12}\right)\nonumber\\[10pt]
&=&\frac{108}{(N^2_c-1)3^3}\int_{\rho}\int d\phi_{1}d\phi_{2}d\phi_{3}e^{in\left(\phi_{1}+\phi_{2}-2\phi_{3}\right)}\cos\left(\mathbf{k}_{1}\cdot\Delta\mathbf{r}_{12}\right)\cos\left(\mathbf{k}_{2}\cdot\Delta\mathbf{r}_{12}\right)\cos\left(\mathbf{k}_{3}\cdot\Delta\mathbf{r}_{12}\right),\nonumber\\[10pt]
\label{dt1}
\end{eqnarray}
where $ac^{3,2}$ is the dipole contribution into the cumulant.
Below we shall denote the asymmetric cumulant for tripole as $T_n$ and the asymmetric cumulant for 3-gluon dipole as $\tilde{T}_n$, while reserving the notation $ac_n$ for total asymmetric three point cumulant for the case of general $N,m$.
The total value of the three point cumulant
\beq
ac_n=ac^{3,2}_n+ac^{3,3}_n.
\eeq
\subsection{Simplifying $T_n$.}
We start by defining $\alpha_{ij}$ to be the azimuthal phase of $\overrightarrow{r}_{ij}\equiv\overrightarrow{r}_{i}-\overrightarrow{r}_{j}$. 

\par We can now take the integral over the 3 azimuthal angles of vectors $\vec k_1,\vec k_2,\vec k_3$ using:
\beq
\int_{0}^{2\pi}d\phi_{1}e^{in\phi_{1}}\cos\left(k_{1}\Delta r_{12}\cos\left(\phi_{1}-\alpha_{12}\right)\right)=\pi i^{n}e^{in\alpha_{12}}\left(1+\left(-1\right)^{n}\right)J_{n}\left(k_{1}\Delta r_{12}\right)\label{bessel}
\eeq
Here  $J_{n}\left(z\right)$ is the $n$-th Bessel function of the first kind. Using  Eq. \ref{bessel} we obtain 
\beq
T_n(k_1,k_2,k_3)=\frac{2^{4}\pi^{3}2\left(1+\left(-1\right)^{n}\right)^{2}}{3\left(2\pi\right)^{3}}\int_{\rho}e^{in\left(\alpha_{12}+\alpha_{23}-2\alpha_{31}\right)}J_{n}\left(k_{1}\left|\mathbf{r}_{12}\right|\right)J_{n}\left(k_{2}\left|\mathbf{r}_{23}\right|\right)J_{2n}\left(k_{3}\left|\mathbf{r}_{31}\right|\right).
\eeq
Note that due to antisymmetry $\vec k\rightarrow -\vec k$ all correlation functions  with odd $n$'s vanish.
\par Consider now  the integral over the sources:
\beq
\left(\begin{array}{ccc}\overrightarrow{r}_{1} & \overrightarrow{r}_{2} & \overrightarrow{r}_{3}\end{array}\right)\rightarrow\left(\begin{array}{ccc}\overrightarrow{r}_{12} & \overrightarrow{r}_{23} & \overrightarrow{r}_{3}\end{array}\right)=\left(\begin{array}{ccc}\overrightarrow{r}_{1}-\overrightarrow{r}_{2} & \overrightarrow{r}_{2}-\overrightarrow{r}_{3} & \overrightarrow{r}_{3}\end{array}\right).
\eeq
We can simplify the latter expression since  the integral over $d^2r_3$ is a simple gaussian integral:
\beq
\int d^{2}r_{3}\frac{e^{-\frac{r_{1}^{2}+r_{2}^{2}+r_{3}^{2}}{2B}}}{\left(2\pi B\right)^{3}}=\int d^{2}r_{3}\frac{e^{-\frac{\left(\overrightarrow{r}_{12}+\overrightarrow{r}_{23}+\overrightarrow{r}_{3}\right)^{2}+\left(\overrightarrow{r}_{23}+\overrightarrow{r}_{3}\right)^{2}+r_{3}^{2}}{2B}}}{\left(2\pi B\right)^{3}}=\frac{e^{-\frac{r_{12}^{2}+\overrightarrow{r}_{12}\cdot\overrightarrow{r}_{23}+r_{23}^{2}}{3B}}}{3\left(2\pi B\right)^{2}}.\label{gauss}
\eeq
Using Eq. (\ref{gauss})  we obtain the final expression for $T_n$:
\begin{eqnarray}
T_{n}&\equiv&\frac{2^{4}}{3^2\left(2\pi B\right)^{2}}\int dr_{12}dr_{23}d\alpha_{12}d\alpha_{23}r_{12}r_{23}e^{-\frac{r_{12}^{2}+r_{12}r_{23}\cos\left(\alpha_{12}-\alpha_{23}\right)+r_{23}^{2}}{3B}+in\left(\alpha_{12}+\alpha_{23}\right)}\nonumber\\[10pt]
&\times&\left(\frac{r_{12}\cos\left(\alpha_{12}\right)+r_{23}\cos\left(\alpha_{23}\right)-i\left(r_{12}\sin\left(\alpha_{12}\right)+r_{23}\sin\left(\alpha_{23}\right)\right)}{\sqrt{r_{12}^{2}+2r_{12}r_{23}\cos\left(\alpha_{12}-\alpha_{23}\right)+r_{23}^{2}}}\right)^{2n}\nonumber\\[10pt]
&\times& J_{n}\left(k_{1}r_{12}\right)J_{n}\left(k_{2}r_{23}\right)J_{2n}\left(k_{3}\sqrt{r_{12}^{2}+2r_{12}r_{23}\cos\left(\alpha_{12}-\alpha_{23}\right)+r_{23}^{2}}\right).\nonumber\\[10pt]
\label{last}
\end{eqnarray}
Where we use $e^{in\alpha}=\left(\frac{x+iy}{\left|\overrightarrow{r}\right|}\right)^{n}$ to find $\alpha_{31}$ in terms of $\alpha_{12},\alpha_{23}$ ( Here $\vec r=(x,y)$, hence $x+iy=r\exp(i\alpha)$).
\par For very small momenta  ( i.e. all $k_i<< 1/B^{1/2}$).
we obtain 
\beq
T_n\simeq \frac{2^{4-2n}\left(B^2Sym(k_{1}k_{2}k_{3}^{2})\right)^{n}}{3\left(n!\right)^{2}}.
\label{sit1a}
\eeq
where $Sym$ means symmetrization over 3 gluons (and division by 1/3). The details of calculation are given in Appendix A.
\subsection{Simplifying $\tilde{T}_n$}
In the same way as we did to $T_n$ we can get a simplified  form of $\tilde{T}_n$, taking the integral over the phases gives us:
\begin{eqnarray}
\tilde{T}_{n}(k_1,k_2,k_3) &=&\frac{2^{2}\cdot2\left(1+\left(-1\right)^{n}\right)^{2}}{\left(2\pi\right)^{3}}\int_{\rho}e^{in\left(\alpha_{12}+\alpha_{12}-2\alpha_{12}\right)}J_{n}\left(k_{1}r_{12}\right)J_{n}\left(k_{2}r_{12}\right)J_{2n}\left(k_{3}r_{12}\right)\nonumber\\[10pt]
&=&\frac{2^{4}\cdot2\left(1+\left(-1\right)^{n}\right)^{2}}{\left(2\pi\right)^{3}}\int_{\rho}J_{n}\left(k_{1}r_{12}\right)J_{n}\left(k_{2}r_{12}\right)J_{2n}\left(k_{3}r_{12}\right).
\end{eqnarray}
We note that here too due to antisymmetry $\vec k\rightarrow -\vec k$ all correlation functions  with odd $n$'s vanish.
The integral over the position of the sources is now a gaussian with two vector variables. Using the transformation:
\beq
\left(\begin{array}{cc}\overrightarrow{r}_{1} & \overrightarrow{r}_{2}\end{array}\right)\rightarrow\left(\begin{array}{cc}\overrightarrow{r}_{12} & \overrightarrow{r}_{2}\end{array}\right)=\left(\begin{array}{cc}\overrightarrow{r}_{1}-\overrightarrow{r}_{2} & \overrightarrow{r}_{2}\end{array}\right),
\eeq
we can take the integral over $d^2r_2$ and over the azimuthal part of $\overrightarrow{r}_{12}$:
\beq
\int d^{2}r_{2}d\alpha_{12}\frac{e^{-\frac{r_{1}^{2}+r_{2}^{2}}{2B}}}{\left(2\pi B\right)^{2}}=\int d^{2}r_{2}d\alpha_{12}\frac{e^{-\frac{\left(\overrightarrow{r}_{12}+\overrightarrow{r}_{2}\right)^{2}+r_{2}^{2}}{2B}}}{\left(2\pi B\right)^{2}}=\frac{e^{-\frac{r_{12}^{2}}{4B}}}{2B}.\label{gauss2}
\eeq
Using Eq. (\ref{gauss2})  we obtain the final expression for $\tilde{T}_n$:
\beq
\tilde{T}_{n}\equiv\frac{2}{B}\int dr_{12}r_{12}e^{-\frac{r_{12}^{2}}{4B}}J_{n}\left(k_{1}r_{12}\right)J_{n}\left(k_{2}r_{12}\right)J_{2n}\left(k_{3}r_{12}\right).
\label{lasttilde}
\eeq
\par For very small momenta  ( i.e. all $k_i << 1/B^{1/2}$).
we obtain 
\beq
\tilde{T}_n\simeq \frac{2^{2}\left(B^{2}Sym\left(k_{1}k_{2}k_{3}^{2}\right)\right)^{n}}{\left(n!\right)^{2}}.
\label{sit1b}
\eeq
where $Sym$ means symmetrization over 3 gluons (and division by 1/3). The details of calculation are given in Appendix A.

\subsection{Numerical Results.}
\par The value of  three point cumulant is 
\beq
ac_n=T_n/(N^2_c-1)^2+\tilde T_n/(N^2_c-1)
\eeq
where n is the harmonics number ( we depict n=2,4 cases).
\par \par We cannot calculate the  integrals $T_n$ and $\tilde{T}_n$ analytically, so we will depict several types of different behaviour of $T_2, T_4, \tilde{T}_2$ and $\tilde{T}_4$ as  functions of momenta.  On the other hand the correlators
have nontrivial structure as functions of $k_1,k_2,k_3$. Namely they depend in the polar coordinates:
\beq k_3=k_r\cos (\theta)
 \,\,k_1=k_r\sin(\theta)\cos(\phi)\,\, k_2=k_r\sin(\theta)\sin(\phi)\,\, k_r=\sqrt{k_1^2+k_2^2+k_3^2}
\eeq
in a nontrivial way: the value of the cumulant depends not ony on $k_r$ but also on $\theta,\phi$.
Indeed, already for very small $k_1,k_2,k_3 \rightarrow 0$
\beq
ac^3_2\sim k_1^2k_2^2k_3^2(k_1^2+k_2^2+k_3^2)\sim k_r^8\sin(2\theta)^2\sin(2\phi)^2\sin(\theta )^2
\eeq
\par In order to better
 understand the structure of the cumulant we shall  consider 3 cases:
\begin{itemize}
 \item{\bf Case I}: All the momenta are equal to each other, $k_1=k_2=k_3=k$.
 \item{\bf Case II}: $\theta,\phi=const$, and we consider the cumulant and its parts as functions of $k_r$
 \item{\bf Case III}: We consider the cumulant as function of $\theta,\phi$ for several values of $k_r$.
\end{itemize}
\begin{figurehere}
\begin{center}
\includegraphics[scale=0.35]{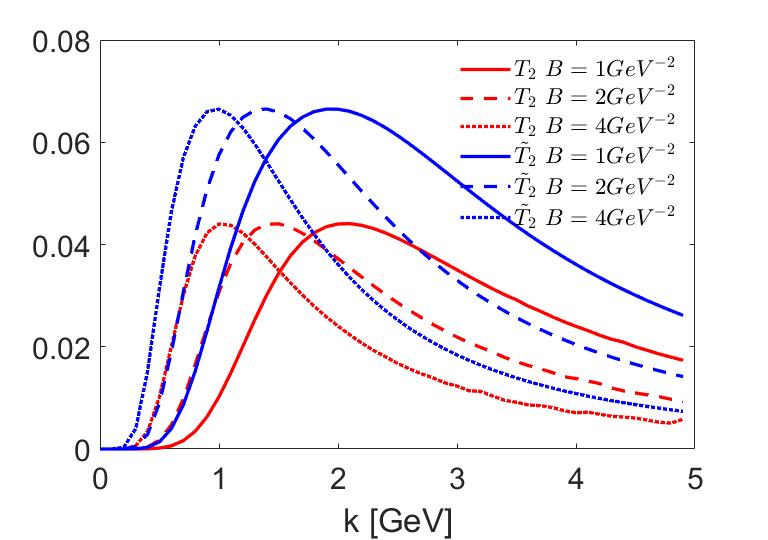}\hspace{0.2cm}
\includegraphics[scale=0.35]{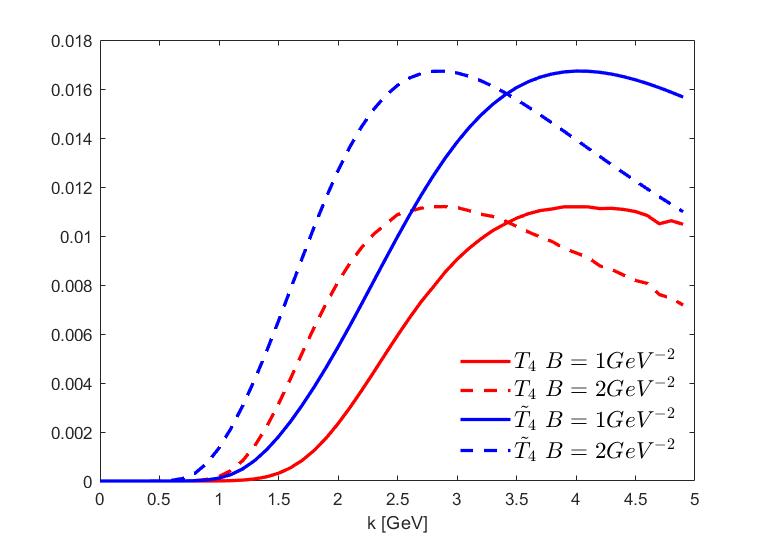}\hspace{0.2cm}
\label{Fig3}
\caption{Above: The integrals $T_2$ and $\tilde{T}_2$  when all momenta are equal, $k_1=k_2=k_3=k$ for different values of the parameter $B=1(2,4)GeV^{-2}$ in full (dashed,dotted) line.
 Below: The integrals  $T_4$ and $\tilde{T}_4$  when all momenta are equal, $k_1=k_2=k_3=k$ for different values of the parameter $B=1(2)GeV^{-2}$ in full (dashed) line. }
\end{center}
\end{figurehere}
\begin{figurehere}
\begin{center}
\includegraphics[scale=0.35]{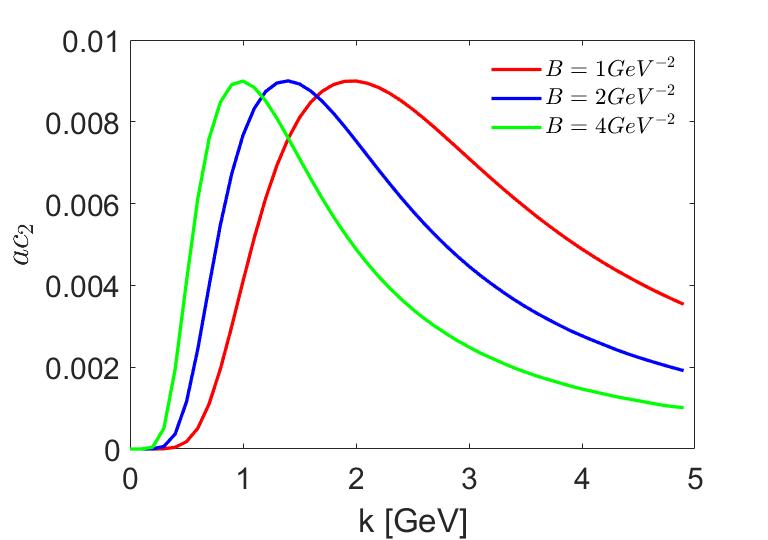}\hspace{0.5cm}
\label{Fig3a }
\caption{The cumulant $ac_2$ for  $N=m=3, N_c=3$, $k_1=k_2=k_3$ }
\end{center}
\end{figurehere}
\begin{figurehere}
\begin{center}
\includegraphics[scale=0.35]{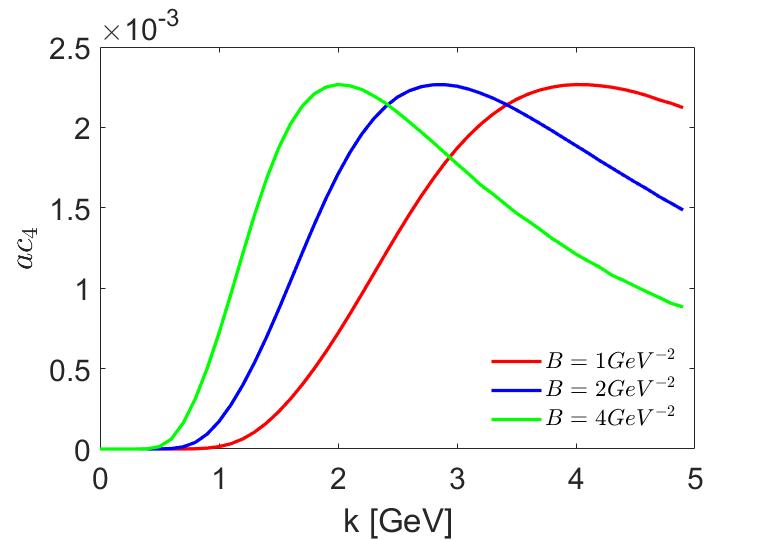}\hspace{0.5cm}
\label{Fig3c}
\caption{The cumulant $ac_4$ for $N=m=3, N_c=3$ for $k_1=k_2=k_3$}
\end{center}
\end{figurehere}
\par In Fig. 3 above we depict  the dependence of $T_2$  and $\tilde{T}_2$ on $k$. We see that for equal $k$, the integrals increases up to $k$ of order $1/\sqrt{B}$, and then slowly decreases, with value at maximum of order  $T_2\approx  0.04 ,\tilde{T}_2\approx 0.07$ that depend on B very weakly. Note also that the maximum is located approximately at the same place where the maximum for the similar graph for symmetric cumulant  (second harmonic) is located, and their k dependence look very similar.
\par We see that both integrals vanish at $k=0$, and as  $k\rightarrow \infty$ we get slowly $T_2\rightarrow 0$ and $\tilde{T}_2\rightarrow 0$. Both $T_2$ and $\tilde T_2$ have maximum in the same point, and their k-dependence is very similar.
\par We  observe very similar  behaviour as a function of k for  $T_4$ and  $\tilde{T}_4$ in Fig.3 below,
except the fourth harmonics is 4-5 times smaller and the maximum is shifted to larger k-s.
\par In Fig 4 we depict the corresponding full three point cumulant $ac_2$, i.e. the second harmonic dependence on k,
for the model case N=m=3. In this case even for $N_c=3$ the diagram with 3 gluons in a dipole is a dominant one,
giving 90 percent of the value of the cumulant. Similarly we depict $ac_4$ in Fig. 5.
\par An interesting feature of the momentum dependence of the cumulant is that the direction in ($\phi, \theta$ ) along which the cumulant is maximal is the one corresponding to $k_1 = k_2 = k_3$. We illustrate this by considering the dependence of $ac_2$, as well as tripole and dipole diagrams on the direction in k-space in Figs. 6,7.
\begin{figurehere}
\begin{center}
\includegraphics[scale=0.27,angle =270]{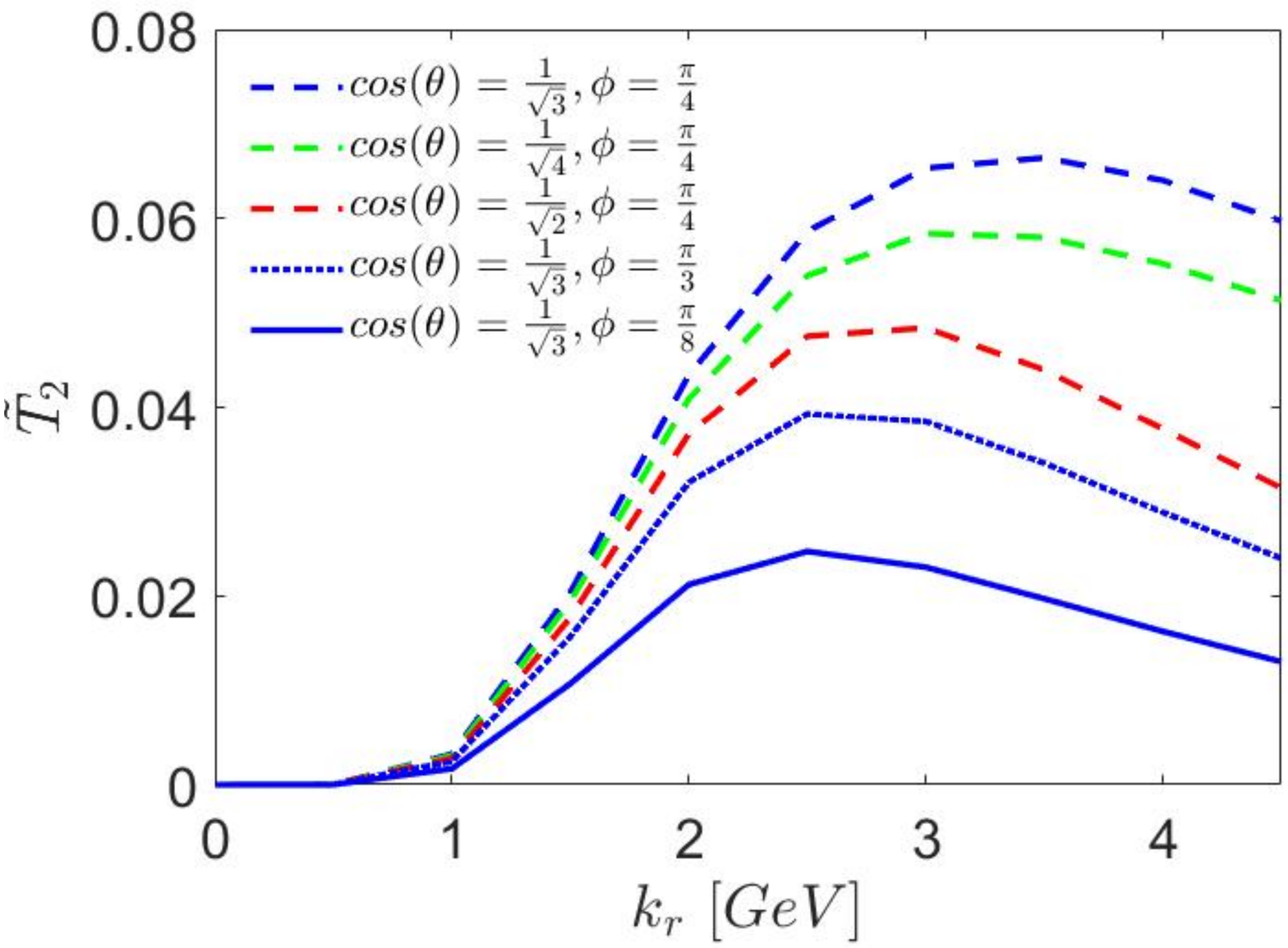}
\includegraphics[scale=0.27,angle=270]{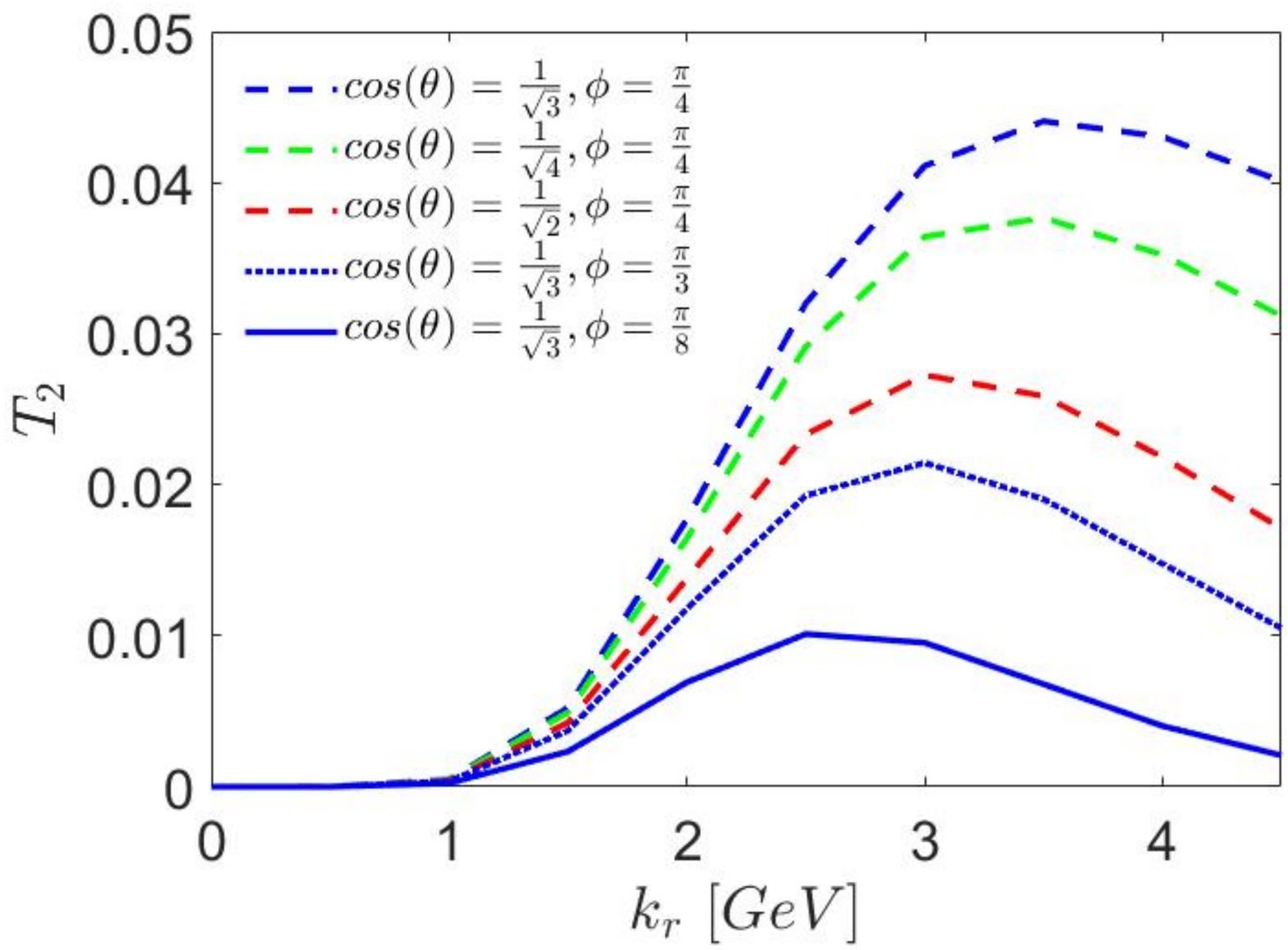}
\label{Fig4}
\caption{The integrals $T_2$  and $\tilde{T}_2$ for different directions, $\theta,\phi=const$, $k_r$ is changing}
\end{center}
\end{figurehere}
\begin{figurehere}
\begin{center}
\includegraphics[scale=0.35,angle =270]{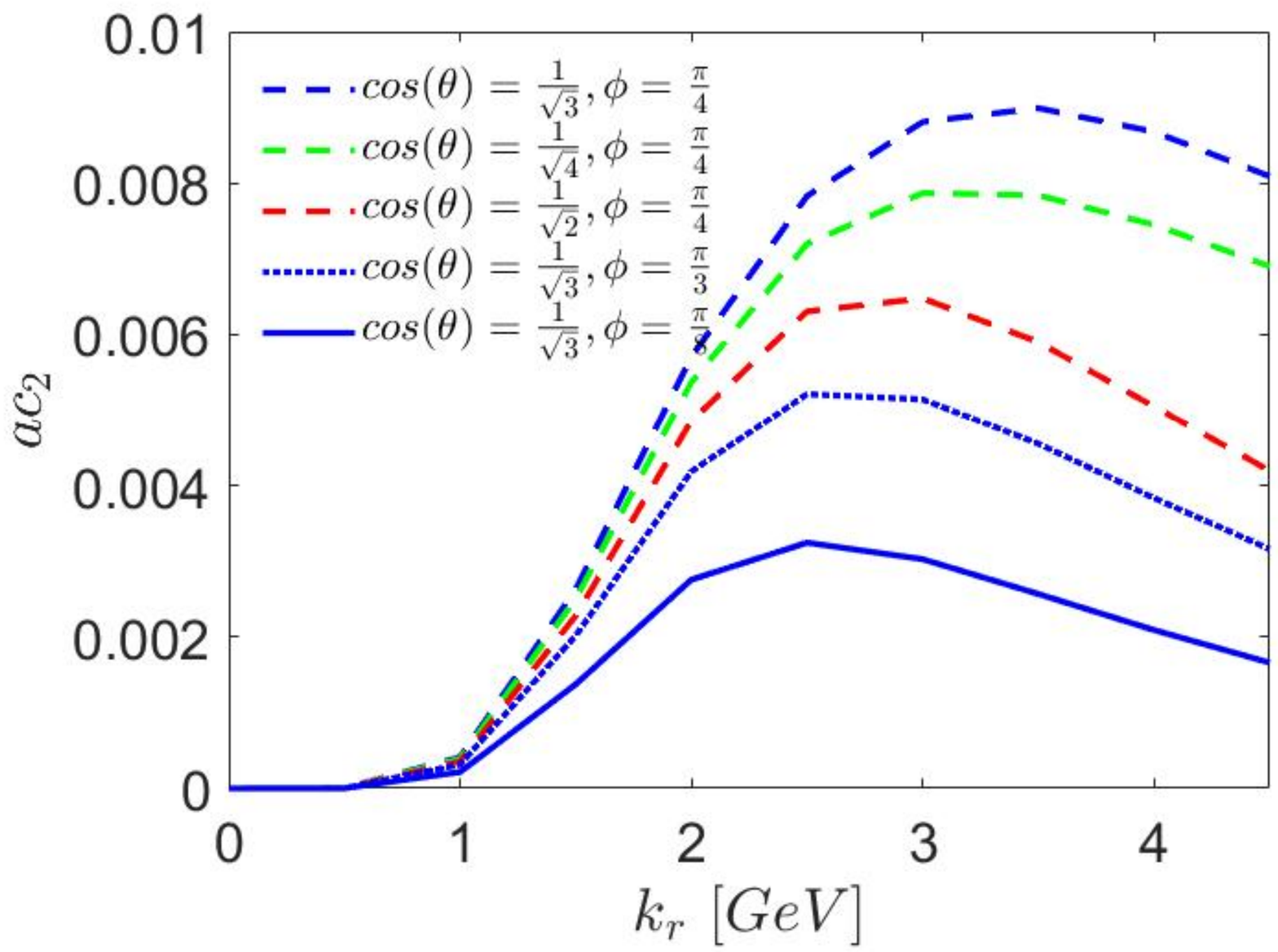}\hspace{0.5cm}
\label{Fig4a}
\caption{ The full cumulant  $ac_2$ for different directions, $\theta,\phi=const$, $k_r$ is changing. B=1 $GeV^{-2}$ }
\end{center}
\end{figurehere}
We see from Figs. 6,7 that indeed  the direction $k_1=k_2=k_3$  (i.e. $\cos(\theta)=1/\sqrt{3},\phi=\pi/4$) corresponds to an absolute maximum. On the other hand 
we see common structure in each direction, with increase up to some maximum value and then slow decrease
depending on the direction.
The Figs. 6,7  are done for $B=1 $GeV$^{-2}$, for other values of B the behaviour is qualitatively similar.

  \par In order to understand better the two dimensional structure we also consider the behaviour 
  of the cumulant as a function of $\theta,\phi$ for $k_r=2,4,6$  GeV. This  is depicted in Figs. 8,9. 

\begin{figurehere}
\begin{center}
\includegraphics[scale=0.27,angle =270]{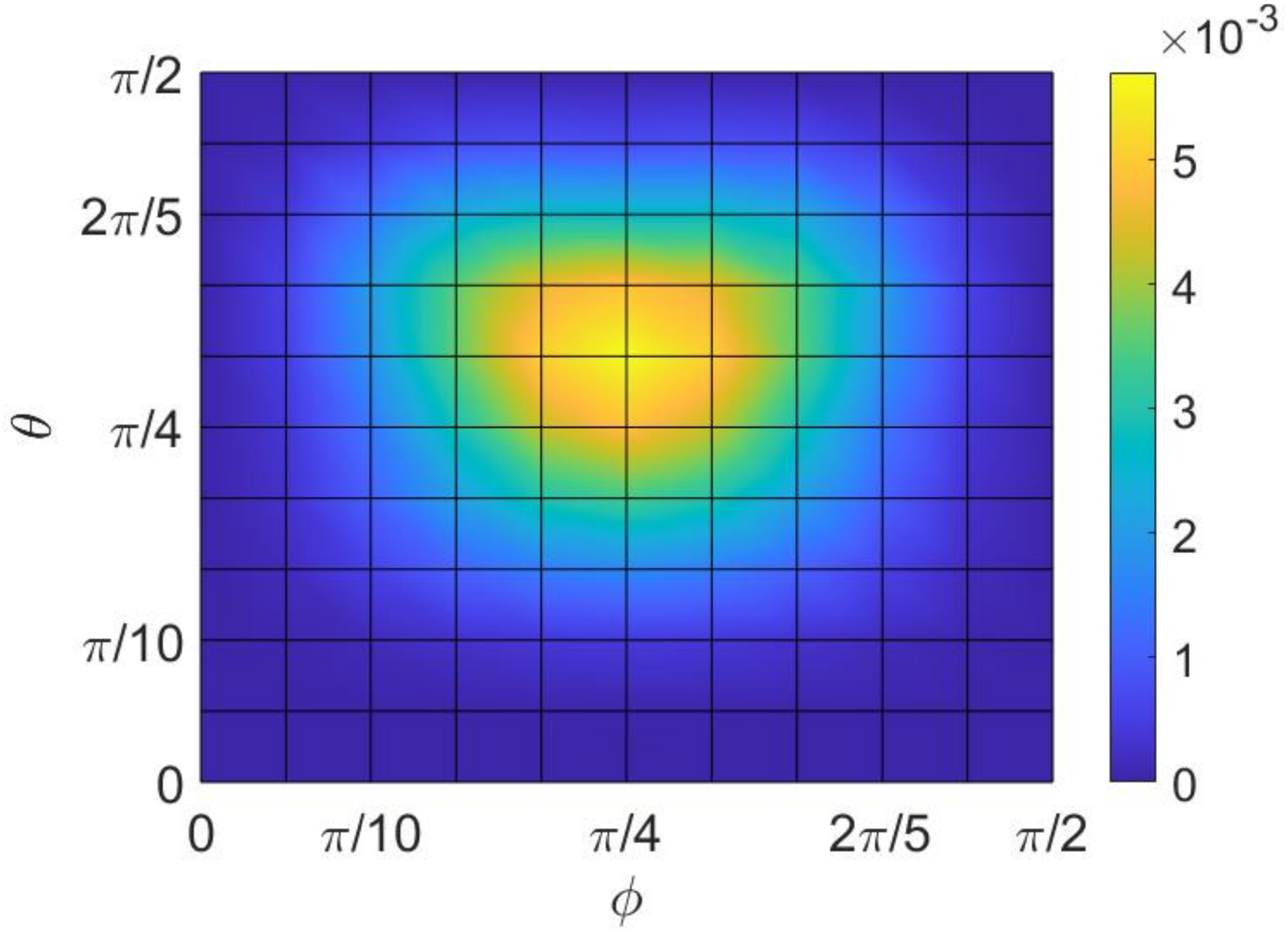}
\includegraphics[scale=0.27,angle =270]{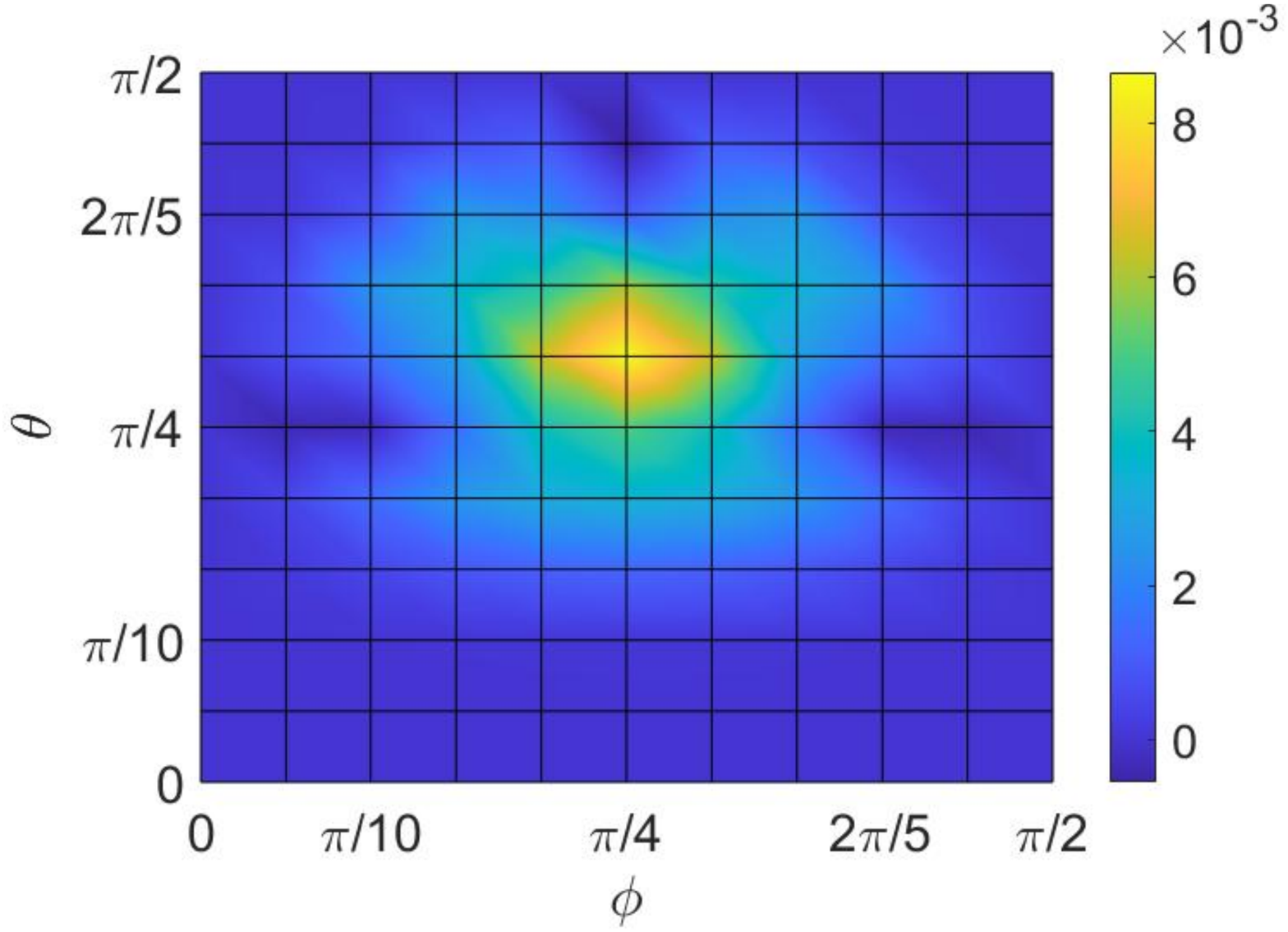}
\includegraphics[scale=0.27,angle =270]{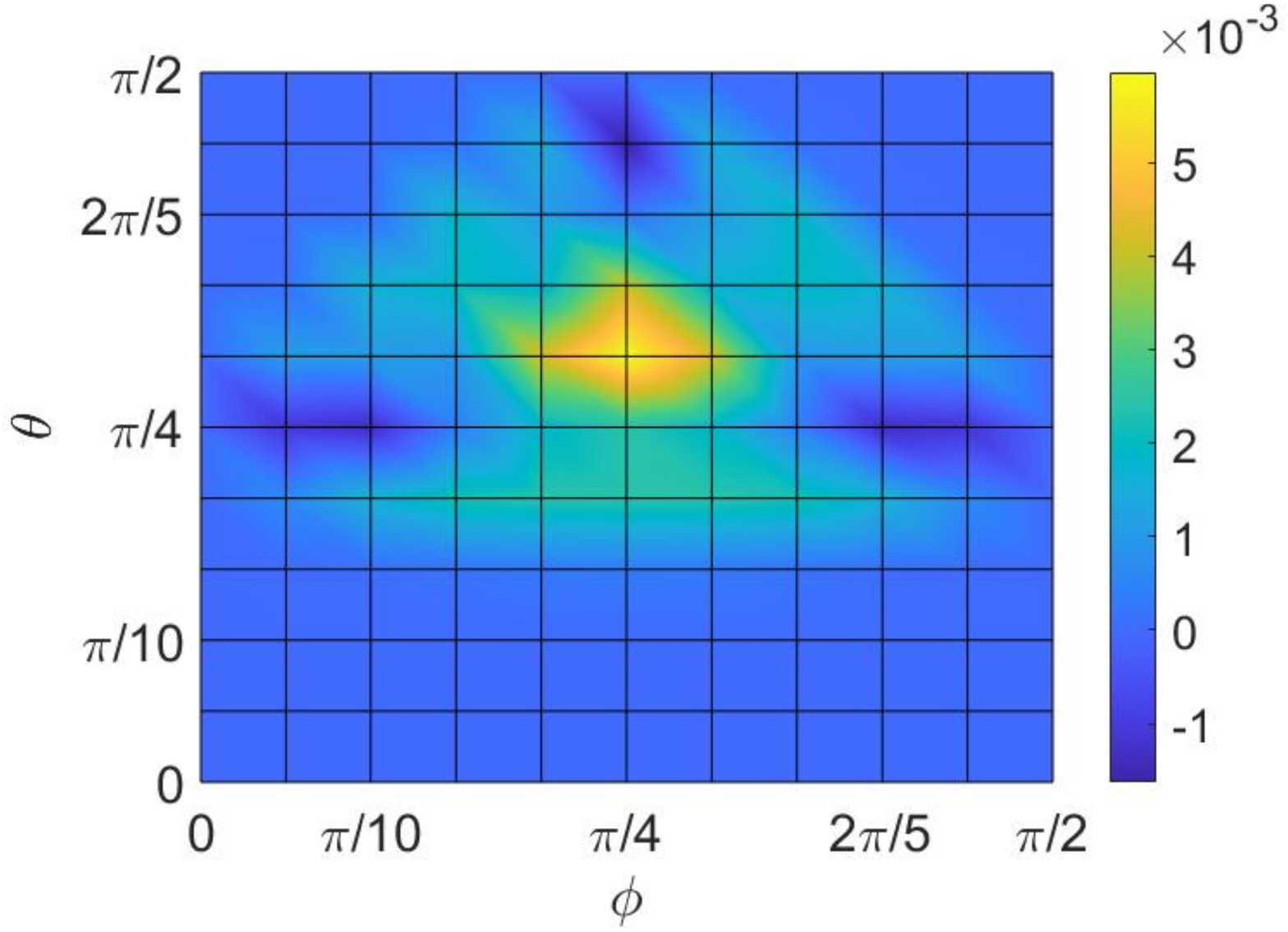}
\label{Fig5b2}
\caption{The cumulant $ac_2$ for N=m=3 $N_c=3$  when $k_r=2,4,6$ GeV and B=1 GeV$^{-2}$}
\end{center}
\end{figurehere}
\begin{figurehere}
\begin{center}
\includegraphics[scale=0.27,angle=270]{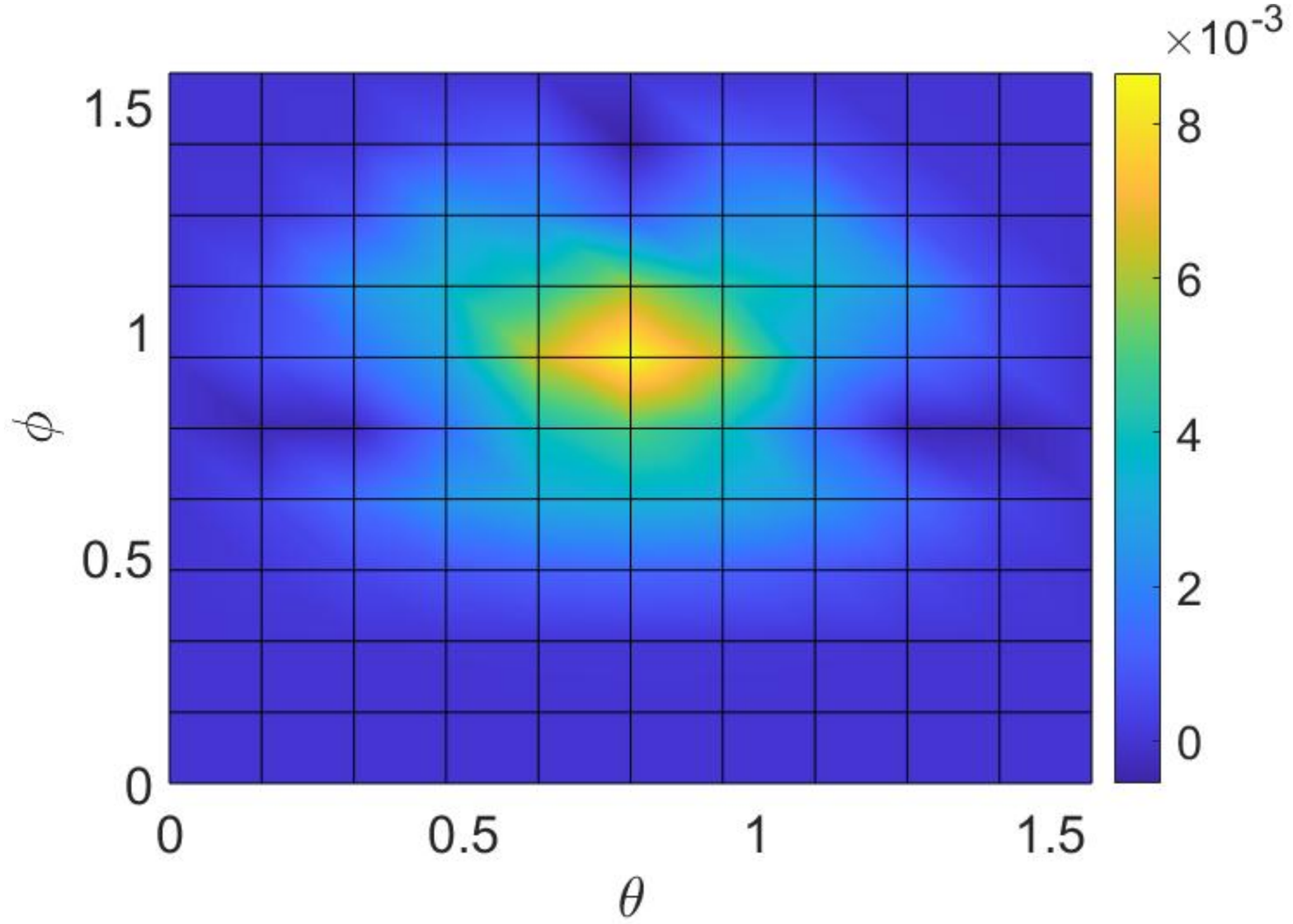}
\includegraphics[scale=0.27,angle =270]{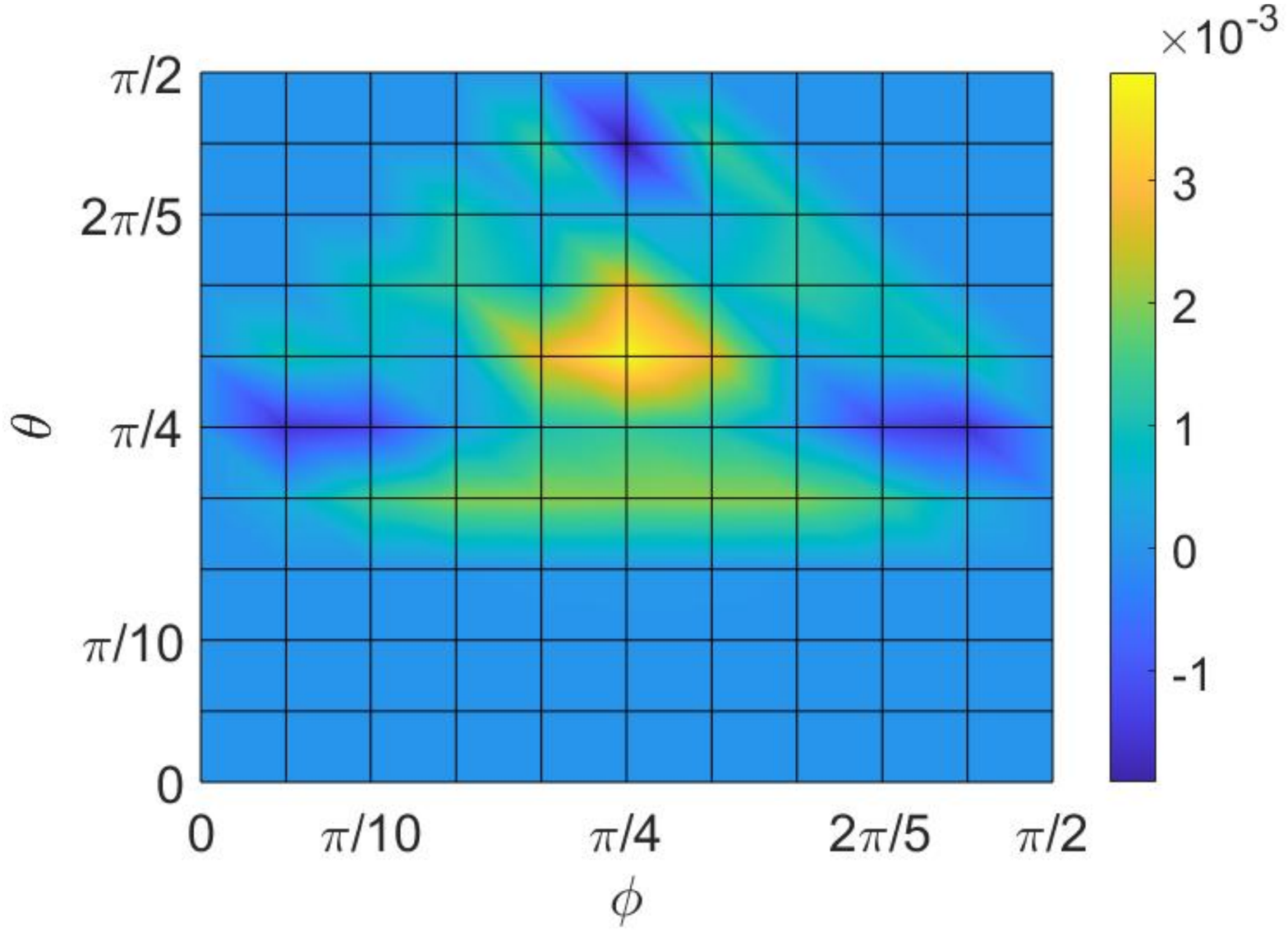}
\includegraphics[scale=0.27,angle =270]{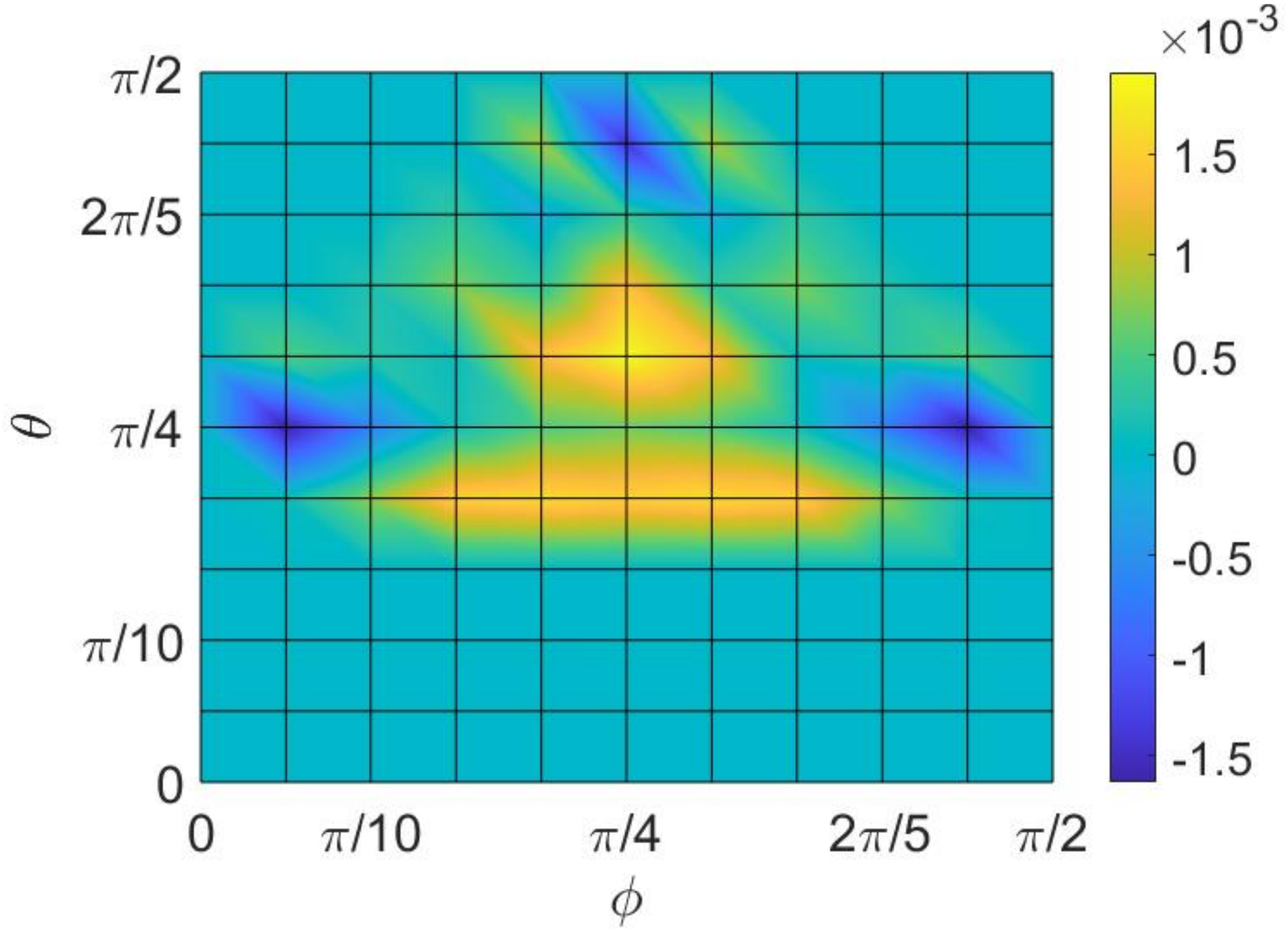}
\label{Fig5c2}
\caption{The cumulant $ac_2$ for N=m=3 $N_c=3$  when $k_r=2,4,6$ GeV and B=4 GeV$^{-2}$}
\end{center}
\end{figurehere}
\par Recall, that the case $B=1$ GeV $^{-2}$corresponds to the best fit 
for  even symmetric cumulants in \cite{15}, while B=4 GeV$^{-2}$ corresponds to the mean field approach
with the effective cross section for MPI two times bigger than the experimental one.  We limit ourselves by depicting  the second harmonic. 
 Note that 
the structure of the cumulant for $B=2,3 $ GeV$^{-2}$ is very similar to the one for B=1 GeV$^{-2}$, so we do not depict it here.
and starts to change only for larger B that correspond to the effective DPS cross sections bigger than the experimental one.

\par Let us note that due to very similar form of the $k-$dependence of $T_n$ and $\tilde T_n$, except their scale,
the similar dependence on k-s will continue for the case of arbitrary  $m,N$ (with different overall coefficient depending on N,m).
\par We also compared  the numerical results for small momenta  with the the analytic expression (\ref{sit1a}) and expression (\ref{sit1b}) and found that they coincide.

\section{\bf High Multiplicity.}
\subsection{Higher Order Diagrams.}
\par We now  consider the general case of $N,m>3$.
It was already noted above that the leading behaviour in powers of $1/(N^2_c-1)$ (and a  resummation of series in terms of $m^2/(N^2_c-1)$) corresponds to diagrams with one tripole and arbitrary number of nonintersecting dipoles.  However in this
case there is also the term contributing to the  cumulant that is suppressed by 1/N but of the first order in 
$1/(N^2_c-1)$. The corresponding resummation, analogous to the resummation for tripole,   will lead to inclusion of the series 
corresponding to the diagrams with one dipole with 3 off-diagonal gluons and up to $[N/2-1]$ nonintersecting dipoles,
such that each of the sources has only 2  (or zero) gluons coming out. Numerically for $N_c=3$
the term  with 3 gluon dipole is a dominant one up to very large multiplicities of order 100.

\par Consequently there are three types of diagrams we  have to consider:
\begin{itemize}
 \item{\bf Type a}: The diagrams with arbitrary (up to N/2)  number of dipoles with 2  off-diagonal gluons. These diagrams 
 were considered in detail in
 \cite{16}. Such diagrams,  contribute to the total cross section $\hat\sigma$.
 \item{\bf Type b}: There is one tripole and $d<N/2$  nonintersecting dipoles with 2 off-diagonal gluons.
 \item{\bf Type c}: There is one  dipole with 3 off-diagonal gluons and $d$  nonintersecting dipoles with 2 off-diagonal gluons each.
\end{itemize}
\par  Let us recall the calculation   of the  combinatorial coefficients
for the case a).We first   pick 2  gluons and 2 sources for each 2-gluon dipole.
 From the picked gluons we get a factor of ${1\choose 2}$:
 \beq
{m\choose 2}{m-2\choose 2}...{m-2d+2\choose 2}\frac{1}{d!}=\frac{\left(m\right)!}{2^{d}d!\left(m-2d\right)!}.\eeq
The sources give us a similar factor of:
 \beq
{N\choose 2}{N-2\choose 2}...{N-2d+2\choose 2}=\frac{\left(N\right)!}{2^{d}\left(N-2d\right)!}.\eeq
\par For type b diagrams we  have an additional multiplier coming from the number of choices 
 of the 3-gluon dipole given by:
\beq 3{m\choose 3}3!{N\choose 2}\eeq
\par For type c the  number of choices  for the tripole give us a factor:
\beq   3{m\choose 3}3!{N\choose 3}\eeq
\par Only diagrams of type b and c contribute to harmonics of $n>0$ so we can write:
\begin{eqnarray}
\frac{d^{3}\hat\sigma}{\prod_{i=1}^{i=3}d\Gamma_{i}}&\propto& N_{c}^{m}(N_{c}^{2}-1)^{N}N^{m}
\prod_{i=1}^{i=3}\vert \overrightarrow{f}(\boldsymbol{k}_{i})\vert^{2}\nonumber\\[10pt]
&\times&\left\{\right.\sum_{d=0}^{\left\lfloor(N-2)/2\right\rfloor}\left(\frac{\hat{D}_{0}F_{corr}^{(2)}\left(N,m\right)}{N^2(N_{c}^{2}-1)}\right)^{d}
\frac{2^33!F_{corr}^{(3)}(N,m)}{N^34(N_{c}^{2}-1)}
\frac{m!N!}{d!3!2!(m-2d-3)!(N-2d-2)!} \nonumber\\[10pt]
&\times&\int_{\rho}\cos\left(\mathbf{k}_{1}\cdot\mathbf{r}_{12}\right)\cos\left(\mathbf{k}_{2}\cdot\mathbf{r}_{12}\right)\cos\left(\mathbf{k}_{3}\cdot\mathbf{r}_{12}\right) \nonumber\\[10pt]
&+&\sum_{d=0}^{\left\lfloor(N-3)/2\right\rfloor}\left(\frac{\hat{D}_{0}F_{corr}^{(2)}\left(N,m\right)}{N^2(N_{c}^{2}-1)}\right)^{d}
\frac{2^33!F_{corr}^{(3)}(N,m)}{N^3(N_{c}^{2}-1)^2}
\frac{m!N!}{d!(3!)^2(m-2d-3)!(N-2d-3)!} \nonumber\\[10pt]
&\times&\int_{\rho}\cos\left(\mathbf{k}_{1}\cdot\mathbf{r}_{12}\right)\cos\left(\mathbf{k}_{2}\cdot\mathbf{r}_{23}\right)\cos\left(\mathbf{k}_{3}\cdot\mathbf{r}_{31}\right)\left.\right\} .\nonumber\\[10pt]
\end{eqnarray}
Here we defined the integral $\hat D_0$ corresponding to the off-diagonal dipole component of the wave function of the 
nucleon fully integrated out i.e. integrated both over the source positions and the  momenta of the gluons:
\beq
\hat{D}_{0}\equiv\int\left(r_{12}dr_{12}\right)\left(\prod_{j=1,2}k_{j}dk_{j}\left(\left|\overrightarrow{f}(\mathbf{k}_{j})\right)\right|^{2}\right)\frac{e^{-\frac{r_{12}^{2}}{4B}}}{2B}J_{0}\left(k_{1}r_{12}\right)J_{0}\left(k_{2}r_{12}\right).
\eeq

This integral is determined by a normalized radiation amplitude $\vec f$ and like in \cite{16} can be considered as a free parameter (coinciding with the one in \cite{16}). This integral   is expected to be between 0 and 1 \cite{16}.
 \par As was already noted above, it was shown in \cite{15} that the diagonal gluons contribute to the interference diagrams by renormalizing them,
 i.e. multiplying by factors $F(m,N)$ that can be easily calculated. For renormalisation  factors $F^{(2)}$ and $F^{(3)}$ connected with diagonal gluons for dipole and tripole diagrams relevant for our discussion  we have the explicit
 expressions:
\beq
F^{(3)}(N,m)=\frac{(m-3)!}{m!}(6(m-2N)N^2+6(N-1)^{m-1}N^{3-m}(-2+m+2N)).
\label{f3}
\eeq
In the limit $N\rightarrow \infty,m\rightarrow \infty, m/N=\bar m =const$, we have 
\beq
F^{(3)}(N,m)\rightarrow F^{(3)}(\bar m)=6\frac{e^{-\bar m}(2+\bar m)}{\bar m^3}+6\frac{\bar m -2}{\bar m^3}.
\eeq
In the same way it was obtained in \cite{15}
\beq
F^{(2)}(N,m)=\frac{2N^{1-m}(N(N-1)^m+mN^m-N^{1+m})}{m(m-1)},
\label{f2}
\eeq
and in the limit  $N\rightarrow \infty,m\rightarrow \infty, m/N=\bar m =const$, we have 
\beq
F^{(2)}(N,m)\rightarrow F^{(2)}(\bar m)=\frac{2\bar m+2e^{-\bar m}-2}{\bar m^2 }.
\eeq
For the the 3-gluon dipole we get a correction factor that is the same as $F^{(3)}(N,m)$, as calculated in appendix B.
For nonintersecting dipoles/tripoles  it is possible to prove that the corresponding renormalisation factors factorize.
\par To find the differential multiplicity we also need to find the total cross section $\hat\sigma$ 
in the same approximation. This cross section is equal to
\begin{eqnarray}
\hat\sigma&\propto& N_{c}^{m}(N_{c}^{2}-1)^{N}N^{m}\nonumber\\[10pt]
&\times&\left\{\right.\left(\sum_{d=0}^{\left\lfloor N/2\right\rfloor}\frac{\hat{D}_{0}F_{corr}^{\left(2\right)}\left(N,m\right)}{N^{2}\left(N_{c}^{2}-1\right)}\right)^{d}\frac{m!N!}{d!\left(m-2d\right)!\left(N-2d\right)!}\nonumber\\[10pt]
&+&\frac{3^{3}F_{corr}^{\left(3\right)}\left(N,m\right)\hat{\tilde{T}}_{0}}{N^{3}\left(N_{c}^{2}-1\right)}\sum_{d=0}^{\left\lfloor(N-2)/2\right\rfloor}\left(\frac{\hat{D}_{0}F_{corr}^{\left(2\right)}\left(N,m\right)}{N^{2}\left(N_{c}^{2}-1\right)}\right)^{d}\frac{m!N!}{d!2!3!\left(m-2d-3\right)!\left(N-2d-2\right)!}\nonumber\\[10pt]
&+&\frac{3^{3}F_{corr}^{\left(3\right)}\left(N,m\right)\hat{T}_{0}}{N^{3}\left(N_{c}^{2}-1\right)^{2}}\sum_{d=0}^{\left\lfloor(N-3)/2\right\rfloor}\left(\frac{\hat{D}_{0}F_{corr}^{\left(2\right)}\left(N,m\right)}{N^{2}\left(N_{c}^{2}-1\right)}\right)^{d}\frac{m!N!}{d!(3!)^2\left(m-2d-3\right)!\left(N-2d-3\right)!}\left.\right\},\nonumber\\[10pt]
\end{eqnarray}
where we define the integrals:
\begin{eqnarray}
\hat{\tilde{T}}_{0}&\equiv&\frac{2}{3^3}\int\left(r_{12}dr_{12}\right)\prod_{j=1}^{3}k_{j}dk_{j}\left|\overrightarrow{f}\left(\mathbf{k}_j\right)\right|^{2}
\frac{e^{-\frac{r_{12}^2}{4B}}}{B}
J_{0}\left(k_{1}r_{12}\right)J_{0}\left(k_{2}r_{12}\right)J_{0}\left(k_{3}r_{12}\right),\nonumber\\[10pt]
\label{Ttil}
\end{eqnarray}
\begin{eqnarray}
\hat{T}_{0}&\equiv&\frac{2^2}{3^5}\int\left(d^2r_{12}d^2r_{23}\right)\prod_{j=1}^{j=3}k_{j}dk_{j}\left|\overrightarrow{f}\left(\mathbf{k}_j\right)\right|^{2}
\frac{e^{-\frac{r_{12}^2+\overrightarrow{r}_{12}\cdot\overrightarrow{r}_{23}+r_{23}^2}{3B}}}{(2\pi B)^2}
\nonumber\\[10pt]
&\times& J_{0}\left(k_{1}r_{12}\right)J_{0}\left(k_{2}r_{23}\right)J_{0}\left(k_{3}\left|\overrightarrow{r}_{12}+\overrightarrow{r}_{23}\right|\right),\nonumber\\[10pt]
\label{T}
\end{eqnarray}
corresponding to the integrated out tripole and integrated out dipole with 3 off-diagonal gluons. Note that the radiation amplitude $\vec f$ defines the $\hat T_0$ and $\hat{\tilde{T}}_{0}$ values, i.e.  they are not free parameter of the model anymore, i.e. if we know $\hat D_0$ the values of $\hat T_0$ and $\hat{\tilde{T}}_{0}$ are correlated with the value of $\hat D_0$.

\par For  the differential multiplicity we obtain:
\begin{eqnarray}
\frac{d^{3}N}{d\Gamma_{1}d\Gamma_{2}d\Gamma_{3}}&=&\frac{d^{3}\sigma}{\sigma d\Gamma_{1}d\Gamma_{2}d\Gamma_{3}}\nonumber\\[10pt]
&\approx&\left(\prod_{i=1}^{i=3}\left|\overrightarrow{f}\left(\boldsymbol{k}_{i}\right)\right|^{2}\right)\hat{\sigma}^{-1}\nonumber\\[10pt]
&\times&\left[\frac{2F_{corr}^{\left(3\right)}\left(N,m\right)}{N^{3}\left(N_{c}^{2}-1\right)}\sum_{d=0}^{\left\lfloor(N-2)/2\right\rfloor}\left(\frac{\hat{D}_{0}F_{corr}^{\left(2\right)}\left(N,m\right)}{N^{2}\left(N_{c}^{2}-1\right)}\right)^{d}\frac{m!N!}{d!2!3!\left(m-2d-3\right)!\left(N-2d-2\right)!}\right.\nonumber\\[10pt]
&\times&\int_{\rho}\cos\left(\mathbf{k}_{1}\cdot\mathbf{r}_{12}\right)\cos\left(\mathbf{k}_{2}\cdot\mathbf{r}_{12}\right)\cos\left(\mathbf{k}_{3}\cdot\mathbf{r}_{12}\right)\nonumber\\[10pt]
&+&\frac{2^{3}F_{corr}^{\left(3\right)}\left(N,m\right)}{N^{3}\left(N_{c}^{2}-1\right)^{2}}\sum_{d=0}^{\left\lfloor(N-3)/2\right\rfloor}\left(\frac{\hat{D}_{0}F_{corr}^{\left(2\right)}\left(N,m\right)}{N^{2}\left(N_{c}^{2}-1\right)}\right)^{d}\frac{m!N!}{d!(3!)^2\left(m-2d-3\right)!\left(N-2d-3\right)!}\nonumber\\[10pt]
&\times&\left.\int_{\rho}\cos\left(\mathbf{k}_{1}\cdot\mathbf{r}_{12}\right)\cos\left(\mathbf{k}_{2}\cdot\mathbf{r}_{23}\right)\cos\left(\mathbf{k}_{3}\cdot\mathbf{r}_{31}\right)\right],
\label{3d}
\end{eqnarray}
where 
\beq
\hat{\sigma}\equiv\sum_{d=0}^{\left\lfloor N/2\right\rfloor}\left(\frac{\hat{D}_{0}F_{corr}^{\left(2\right)}\left(N,m\right)}{N^{2}\left(N_{c}^{2}-1\right)}\right)^{d}\frac{m!N!}{d!\left(m-2d\right)!\left(N-2d\right)!}.
\eeq
\par is the total cross section. We have shown by direct numerical calculation that the contribution to the total cross section of the integrated out 3-gluon dipole and of the integrated out tripole are negligible compered to the 2-gluon dipole terms, so we can ignore the dependence on $\hat{T}_{0}$ and $\hat{\tilde{T}}_{0}$
\par For differential one gluon distribution we have in this approximation:
\beq
\frac{dN}{d\Gamma }=m\left|\overrightarrow{f}\left(\boldsymbol{k}\right)\right|^{2}.
\eeq
\par We will  can now write for the  parts corresponding to  the 3-gluon dipole and a the tripole contribution
\begin{eqnarray}
ac^{3,2}_{n}\left\{ 3\right\}&=&\frac{3^{3}F_{corr}^{(3)}\left(N,m\right)}{{m\choose 3}\hat{\sigma}N^{3}(N_c^2-1)}\tilde{T}_{n}/3\nonumber\\[10pt]
&\times&\sum^{\left\lfloor(N-2)/2\right\rfloor}_{d=0}\left(\frac{\hat{D}_{0}F_{corr}^{\left(2\right)}(N,m)}{N^{2}\left(N_{c}^{2}-1\right)}\right)^{d}
\frac{m!N!}{d!2!3!(m-2d-3)!(N-2d-2)!}\nonumber\\[10pt]
ac^{3,3}_{n}\left\{ 3\right\}&=&\frac{3^{3}F_{corr}^{(3)}\left(N,m\right)}{{m\choose 3}\hat{\sigma}N^{3}(N_c^2-1)^2}T_{n}\nonumber\\[10pt]
&\times&\sum^{\left\lfloor(N-3)/2\right\rfloor}_{d=0}\left(\frac{\hat{D}_{0}F_{corr}^{\left(2\right)}(N,m)}{N^{2}\left(N_{c}^{2}-1\right)}\right)^{d}
\frac{m!N!}{d!(3!)^2(m-2d-3)!(N-2d-3)!}\nonumber\\[10pt]
ac_n&=&ac^{3,2}_{n}\left\{ 3\right\}+ac^{3,3}_{n}\left\{ 3\right\}.
\label{ex}
\end{eqnarray}
\par Note that taking $\hat{D}_0=0$ or equivalently, only $d=0$ term in the expansion for $\hat{\Sigma}$ we return to the result (\ref{dt}) for  for N=m=3 
of the previous section.
\subsection{Numerical Results.} 
\begin{figurehere}
\begin{center}
\includegraphics[scale=0.35]{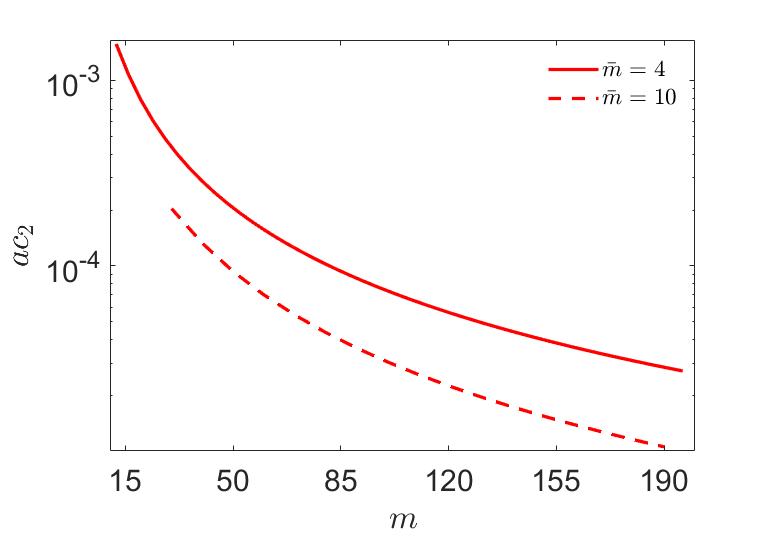}\hspace{3cm}
\label{Fig19}
\caption{The dependence of the maximum value of $ac_2$ (as a function of momenta) on multiplicity  $m$ for different values of $\bar{m}$ where $\hat{D}_0=0.1$ and $N_c=3$.}
\end{center}
\end{figurehere}
\begin{figurehere}
\begin{center}
\includegraphics[scale=0.35]{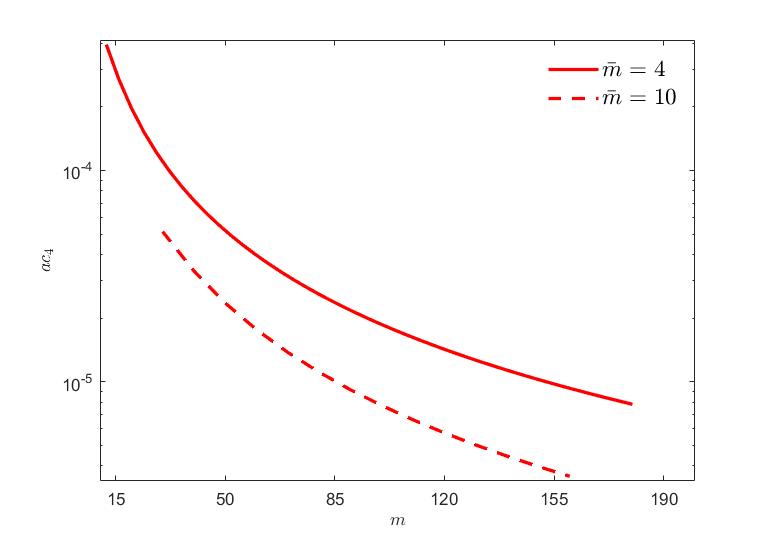}\hspace{3cm}
\label{Fig19a}
\caption{The dependence of the maximum value of $ac_4$ (as a function of momenta) on multiplicity  $m$ for different values of $\bar{m}$ where $\hat{D}_0=0.1$ and $N_c=3$.}
\end{center}
\end{figurehere}
\begin{figurehere}
\begin{center}
\includegraphics[scale=0.35]{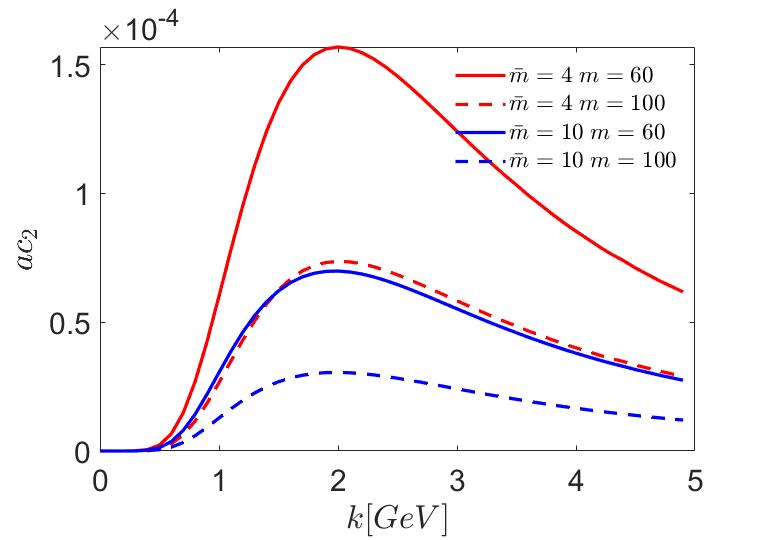}\hspace{3cm}
\label{Fig7b}
\caption{The form of the $ac_2$ for $k_1=k_2=k_3$ and for different values of multiplicity m, with  $\hat{D}_0=0.1$, $B=1$ GeV$^{-2}$,$N_c=3$.}
\end{center}
\end{figurehere}
\par We now look at $ac$ as a function of the multiplicity for different values of $\bar{m}\equiv m/N$. 
\par We first look at $ac_2$  as a function of $m$ for fixed values of  $\bar{m}=4,10$. 
In Fig. 10 we depict the dependence on m of the maximum value of $ac_2$ as function of transverse momenta.
We see that the value of $ac_2$ decreases slowly with multiplicity, and the characteristic scale of $ac_2$ for moderate 
$m\sim 50$ is of order $2\times 10^{-4}$.
\par It is interesting to note that the scale of two point correlator  
$v^2_2\equiv sc_2\left\{2\right\}$,
calculated in \cite{16} was $\sim 4-5\times 10^{-3}$, i.e. we observe a  decrease of order  $2(N^2_c-1)$  going from
$v_2^2$ to  $ac_2$, and the ratio between the two
 very weakly depends on multiplicity. 
\par In Fig 11 we depict the analogous dependence of $ac_4$. We depict the k-dependence of $ac_2$ for various multiplicities in Fig. 12. We see that the k-dependence is practically independent on multiplicity (up to an overall scaling factor),
and is the same as in N=m=3 case.

\section{Comparison with the Experimental Results.}
\par It will be interesting to compare our results with the recent experimental data \cite{17}. In that paper the average of the second harmonic over experimental data was taken with momenta varying in two different kinematic regions $k\in\left[0.3,3\right]GeV$ and $k\in\left[0.5,5\right]GeV$ . 
\par Recall that we can completely separate the dependence of the momenta and $n$, in the form of $\tilde{T}_{n}$ and $T_n$, and the dependence on all other parameters like multiplicity, number of sources, $N_c$ and the model constant $\hat{D}_0$, It is convenient to define:
\begin{eqnarray}
R^{3,2}\left(N,m,N_c,\hat{D}_0\right)&\equiv&\frac{ac^{3,2}_{n}\left\{ 3\right\}}{\tilde{T}_{n}}\nonumber\\[10pt]
R^{3,3}\left(N,m,N_c,\hat{D}_0\right)&\equiv&\frac{ac^{3,3}_{n}\left\{ 3\right\}}{T_{n}}.
\end{eqnarray}
 Since all of the dependence on the momenta in our model is contained in k-dependent functions $T_n(k_1,k_2,k_3)$ and $\tilde{T}_n(k_1,k_2,k_3)$ we can calculate the averages:
 \bigskip
 \begin{center}
\begin{tabular}{ |c|c|c|c| } 
 \hline
 & $B=1$ GeV$^{-2}$ & $B=2$ GeV$^{-2}$&$B=4$ GeV$^{-2}$ \\ 
 \hline
$ \left\langle T_{2}\right\rangle _{0.3-3GeV}=\frac{\int_{0.3}^3dk_{1}dk_{2}dk_{3}T_{2}}{\left(3-0.3\right)^{3}} $& $9.4 \times 10^{-3}$& $6.5 \times 10^{-3}$&$2.7 \times 10^{-3} $\\ 
\hline
$ \left\langle\tilde{T}_{2}\right\rangle _{0.3-3GeV}=\frac{\int_{0.3}^3dk_{1}dk_{2}dk_{3}T_{2}}{\left(3-0.3\right)^{3}} $& $2.1 \times 10^{-2}$& $1.6 \times 10^{-2}$&$1.1 \times 10^{-2} $\\ 
\hline
$ \left\langle T_{2}\right\rangle _{0.5-5GeV}=\frac{\int^5_{0.5}dk_{1}dk_{2}dk_{3}T_{2}}{\left(5-0.5\right)^{3}}$& $4.1 \times 10^{-3}$& $1.4 \times 10^{-3}$&$1.5 \times 10^{-3}$\\ 
\hline
$ \left\langle\tilde{T}_{2}\right\rangle _{0.5-5GeV}=\frac{\int^5_{0.5}dk_{1}dk_{2}dk_{3}T_{2}}{\left(5-0.5\right)^{3}}$& $1.4 \times 10^{-2}$& $8.7 \times 10^{-3}$&$5.0 \times 10^{-3}$\\ 
 \hline
\end{tabular}
\end{center}
\bigskip

\par The value of the cumulant is obtained by calculating:
\beq
\left\langle ac_2\left\{3\right\}\right\rangle=R^{3,2}\left\langle\tilde{T}_{2}\right\rangle+R^{3,3}\left\langle T_{2}\right\rangle .
\eeq
The results are depicted in Fig.13:
\begin{figurehere}
\begin{center}
\includegraphics[scale=0.35]{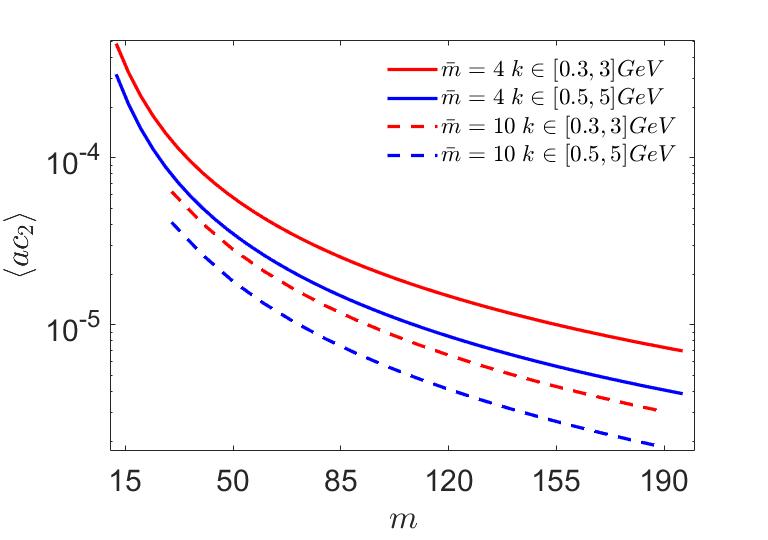}\hspace{0.5cm}
\label{Fig15}
\caption{The 3 point cumulant $ac^3$ averaged over region $0.5<k_i<3$ and $0.5<k_i<5$ (i=1,2,3)}
\end{center}
\end{figurehere}

\par   We depict in Fig. 14 the theoretical value of the second harmonic (integrated over the region  $0.5\le k\le 3$
 together with experimental data, namely the ATLAS result after additional analysis, done in \cite{zhang}
to eliminate nonflow effects. Note that the values of m, that are obtained here using LPHD concept,
correspond to a total number of soft hadrons which is approximately $m\sim 1.5 N_{\rm charged}$, where 
 $N_{\rm charged}$ is a number of charged particles measured in the ATLAS experiment. 
 (The high multiplicity sample used by ATLAS is dominated by  $\pi$ mesons
\cite{zia},  the factor 3/2 then comes from isotopic invariance, since $\pi$ mesons form a triplet in the isotopic space).
 The two different 
 types of data, with or without gap are depicted, meaning the gap of 0.5 units between subevents to limit nonflow
 is taken or not are depicted. We refer the reader to \cite{17,zhang}  for the details of the experimental analysis.
\begin{figurehere}
\begin{center}
\includegraphics[scale=0.35,angle=270]{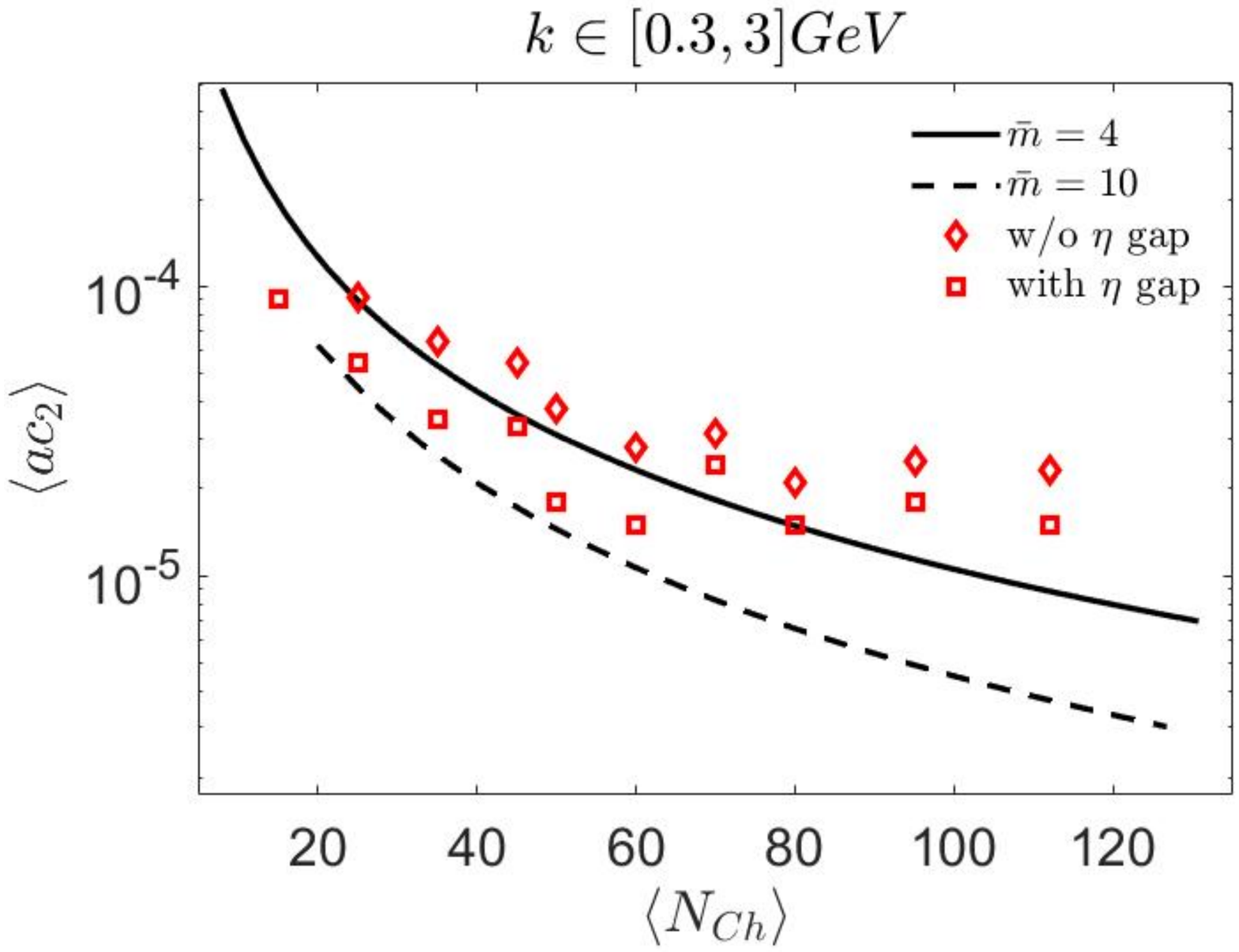}\hspace{0.5cm}
\label{Fig17}
\caption{The 3 point cumulant $ac_2$ averaged over region $0.5<k_i<3$ and $0.5<k_i<3$ (i=1,2,3),
compared with experimental data. The experimental data is depicted after additional analysis \cite{zhang} of the ATLAS results, made to minimize nonflow effects, where the two types of experimental points correspond to analysis with or without the gap of 0.5 units between subevents } 
\end{center}
\end{figurehere}
 \par  We see  rather good agreement with experimental data for $m\le 120 (N_{charged}\le 80)$
 and if we average over
region $0.5\le k\le 3$ GeV.
 \par  However for higher multiplicities the theoretical result decreases with total multiplicity m
rather rapidly, contrary to the experimental data, which shows the independence of $ac$ on multiplicity for large $m$.
In addition the ATLAS data for average $ac_2$  over the region $0.5\le k\le 5$ GeV tend to increase relative to average over   $0.5\le k\le 3$ GeV, while Fig. 13 shows the opposite trend.
Note however that these averages are very sensitive to explicit k-dependence and even small inaccuracy in k-dependence leads to rather large inaccuracy in the average. Moreover, our results may be less accurate for large transverse momenta, where soft gluon approximation is  less accurate. 
\section{Conclusions.}
\par We have studied the influence of the effects of colour interference and colour flow on the three point asymmetric cumulants using the model \cite{15,16}.  
\par We get qualitative agreement of our results for asymmetric correlator with the scale of available experimental data \cite{17}, at least for moderate multiplicities $m\sim 100$. Note that 
only integrated experimental data is available, decreasing the possibility of detailed comparison with the experimental results.
This data  seems however to be very sensitive on precise transverse momenta dependence. Thus the detailed comparison between theoretical and experimental results demands further measurements, in particular the detailed study of transverse
momenta dependence, as it was done already for symmetric correlators.
\par From our side we carried the detailed study of the transverse momenta dependence and characteristic scale of the
correlator.
\par The discrepancy with the experimental data is seen in the  decrease of the $ac$ cumulant with multiplicity $m$ at high multiplicities, the analogous behaviour was also noted for  symmetric cumulants in \cite{16}. The experimental data indicates that cumulants are virtually independent of multiplicity.
On the other hand let us note that there may be significant uncertainties in the experimental data,
related to separation of the flow and nonflow effects \cite{zhang}.
We expect that further
study of the model, in particular inclusion of higher suppressed diagrams (like quadrupole like ones) will improve the dependence on multiplicity both for symmetric and asymmetric  correlators \cite{BS}.

\par {\bf Acknowledgements} This research was supported by  ISF grant number 2025311. The authors thank M. Strikman for reading the manuscript and J.  Jia for discussion on experimental data and pointing the reference \cite{zhang}.
\appendix
\section{Small momentum limit}
\par Consider the case of the very small transverse momenta for a tripole. Looking at very small momenta, $k_{j}\ll B^{-1/2}$ for all 3 momenta we can take the Taylor expansion of the Bessel functions
 \beq J_{n}\left(z\right)\simeq\frac{z^{n}}{n!2^{n}}\eeq
  to find:
\begin{eqnarray}
T_{n}&\simeq&\frac{2^{4\left(1-n\right)}\left(k_{1}k_{2}k_{3}^{2}\right)^{n}}{3^2\left(n!\right)^{2}\left(2n\right)!\left(2\pi B\right)^{2}}\int dr_{12}dr_{23}d\alpha_{12}d\alpha_{23}r_{12}^{1+n}r_{23}^{1+n}e^{-\frac{r_{12}^{2}+\overrightarrow{r}_{12}\cdot\overrightarrow{r}_{23}+r_{23}^{2}}{3B}}\nonumber\\[10pt]
&\times&\left(\cos\left(\alpha_{12}\right)+\sin\left(\alpha_{12}\right)\right)^{n}\left(\cos\left(\alpha_{23}\right)+\sin\left(\alpha_{23}\right)\right)^{n}\nonumber\\[10pt]
&\times&\left(r_{12}\cos\left(\alpha_{12}\right)+r_{23}\cos\left(\alpha_{23}\right)-i\left(r_{12}\sin\left(\alpha_{12}\right)+r_{23}\sin\left(\alpha_{23}\right)\right)\right)^{2n}.\nonumber\\[10pt]
\label{sit}
\end{eqnarray}
Note that from dimensional analysis  for small momenta
 \beq ac_{n}\propto \left(B^{2}k_{1}k_{2}k_{3}^{2}\right)^{n}\eeq
Simplifying the integral (\ref{sit}) we obtain in the limit of small $k_i$:
\begin{eqnarray}
T_{n}&\simeq&\frac{\left(3B\right)^{2}2^{4\left(1-n\right)}\left(3^2B^2k_{1}k_{2}k_{3}^{2}\right)^{n}}{3^{2}\left(n!\right)^{2}\left(2n\right)!\left(2\pi B\right)^{2}}\int_{{R}^{4}}dx_{1}dx_{2}dy_{1}dy_{2}\nonumber\\[10pt]
&\times&\left.\frac{\partial^{n}}{\partial\alpha^{n}}\frac{\partial^{n}}{\partial\beta^{n}}\frac{\partial^{2n}}{\partial\gamma^{2n}}e^{-\left(x_{1}^{2}+x_{1}x_{2}+x_{2}^{2}+y_{1}^{2}+y_{1}y_{2}+y_{2}^{2}\right)+\alpha\left(x_{1}+iy_{1}\right)+\beta\left(x_{2}+iy_{2}\right)+\gamma\left(x_{1}+x_{2}-i\left(y_{1}+y_{2}\right)\right)}\right|_{\alpha=\beta=\gamma=0}\nonumber\\[10pt]
&=&\frac{2^{4\left(1-n\right)}\left(3^2B^2k_{1}k_{2}k_{3}^{2}\right)^{n}}{\left(n!\right)^{2}\left(2n\right)!\left(2\pi\right)^{2}}\left.\frac{\partial^{n}}{\partial\alpha^{n}}\frac{\partial^{n}}{\partial\beta^{n}}\frac{\partial^{2n}}{\partial\gamma^{2n}}\frac{4\pi^{2}}{3}e^{\frac{2}{3}\gamma\left(\alpha+\beta\right)}\right|_{\alpha=\beta=\gamma=0}\nonumber\\[10pt]
&=&\frac{2^{4\left(1-n\right)}\left(3^2B^2k_{1}k_{2}k_{3}^{2}\right)^{n}}{3\left(n!\right)^{2}\left(2n\right)!}\left(2n\right)!\left(\frac{2}{3}\right)^{2n}=\frac{2^{4+2n}\left(B^2k_{1}k_{2}k_{3}^{2}\right)^{n}}{3\left(n!\right)^{2}}.\nonumber\\[10pt]
\label{sit1bn}
\end{eqnarray}
\par Note that all expressions above are understood to be symmetrized over gluons 1,2,3,
i.e. equal to sum $\frac{1}{3}\sum_{1,2,3}$.
\par We can also use the same approximation to find:
\beq
\tilde{T}_{n}\simeq\frac{2^{1-4n}\left(Sym\left(k_{1}k_{2}k_{3}^{2}\right)\right)^{n}}{B\left(n!\right)^{2}\left(2n\right)!}\int dr_{12}r_{12}^{1+4n}e^{-\frac{r_{12}^{2}}{4B}}.
\label{sittil}
\eeq
This lives us with an integral that is easly solved by taking $u=\frac{r_{12}^2}{4B}$ and we get the result:
\beq
\tilde{T}_{n}\simeq\frac{2^{2}\left(B^{2}k_{1}k_{2}k_{3}^{2}\right)^{n}}{\left(n!\right)^{2}}.
\eeq
\section{3-gluon dipole correction}
\par In the same way they found the correction in Ref. \cite{15}, we can find the correction factor for the case for the 3-gluon dipole. We note that if we are are looking at an ordered list of emitted gluons  with only three off
-diagonal gluons that make the 3-gluon dipole we can divide the diagonal gluons into 4 kinds. If the off-diagonal gluons are $(1,2,3)$ the diagonal gluons can be before $1$, between $1$ and $2$, between $2$ and $3$ and after $3$. The ones between $1$ and $2$ and between $2$ and $3$ will give us a factor of $1/2$ if they are on the same sources as the off-diagonal gluons, and  will give us a factor of 1 otherwise.
We first need the number of incoherent diagrams, which  will be:
\begin{eqnarray}
N_{incoh}&=&\sum_{j_{12}=0}^{m-3}\sum_{j_{23}=0}^{m-3-j_{12}}N^{m-3-j_{12}-j_{23}}(m-2-j_{12}-j_{23})\nonumber\\[10pt]
&\times&\sum_{l_{12}=0}^{j_{12}}\left(\begin{array}{c}j_{12}\\l_{12}\end{array}\right)2^{l_{12}}(N-2)^{j_{12}-l_{12}}\nonumber\\[10pt]
&\times&\sum_{l_{23}=0}^{j_{23}}\left(\begin{array}{c}j_{23}\\l_{23}\end{array}\right)2^{l_{23}}(N-2)^{j_{23}-l_{23}}\nonumber\\[10pt]
&=&\frac{m!}{3!(m-3)!}N^{m-3}
\end{eqnarray}
where $j_{ab}$ counts the number of diagonal gluons between $a$ and $b$ and $l_{ab}$ counts how many of them are on the same gluons as the 3-gluons dipole.
\par The correction coefficient  is then calculated by taking into account the factor of $1/2^l$,
coming from identity $T^aT^bT^a=T^b/2$:
\begin{eqnarray}
F^{(3,2)}(N,m)&=&\frac{1}{N_{incoh}}\sum_{j_{12}=0}^{m-3}\sum_{j_{23}=0}^{m-3-j_{12}}N^{m-3-j_{12}-j_{23}}(m-2-j_{12}-j_{23})\nonumber\\[10pt]
&\times&\sum_{l_{12}=0}^{j_{12}}\left(\begin{array}{c}j_{12}\\l_{12}\end{array}\right)2^{l_{12}}(N-2)^{j_{12}-l_{12}}2^{-l_{12}}\nonumber\\[10pt]
&\times&\sum_{l_{23}=0}^{j_{23}}\left(\begin{array}{c}j_{23}\\l_{23}\end{array}\right)2^{l_{23}}(N-2)^{j_{23}-l_{23}}2^{-l_{23}}\nonumber\\[10pt]
&=&\frac{(m-3)!}{m!}(6(m-2N)N^2+6(N-1)^{m-1}N^{3-m}(-2+m+2N)).
\end{eqnarray}
We note that this is exactly $F^{(3)}(N,m)$, as we can see from \cite{15}.

\end{document}